\documentclass[11pt,a4paper]{article}
\usepackage{jheppub_kim}
\usepackage{rotating}
\usepackage{graphicx,epsfig}
\usepackage{amsmath}
\usepackage {amssymb}
\usepackage{subfigure}
\usepackage{multirow}
\usepackage{filecontents,catchfile}
\usepackage{float}
\usepackage{perpage}
\usepackage{mathtools}
\MakeSorted{figure}
\MakeSorted{table}
\usepackage[utf8x]{inputenc}
\usepackage{relsize}
\usepackage{cleveref}
\usepackage{pgfplots}
\usepackage{array,multirow}
\usepackage{soul}
\usepackage{subfigure}
\usepackage{dsfont}
\usepackage{hyperref}
\usepackage{txfonts}
\usepackage{amsfonts}
\usepackage{newlfont}
\usepackage{times}
\usepackage{amssymb}
\usepackage{secdot}
\usepackage{cmap}
\usepackage[explicit]{titlesec }

\graphicspath{ {./images/} }
\titlespacing{\paragraph}{0pt}{0pt}{.5em}[]

\usepackage[font=scriptsize]{caption}

\title{Non-Extensive Transverse Momentum Distribution For Identified Particles at $\sqrt{S_{NN}}= $ 7.7, 11.5, 19.6, 27, 39 \hspace{0.05cm} GeV.}

\author[a]{Wafaa Saleh}
\author[b]{Asmaa G. Shalaby}

\affiliation[a]{Basic Science Department, Faculty of Computer and Information Sciences, Ain Shams University, Cairo 11566, Egypt.}
\affiliation[a]{Physics Department, Faculty of Science, Ain Shams University, Cairo 11566, Egypt.}

\affiliation[b]{Physics Department, Faculty of Science, Benha University, Benha 13518, Egypt.}

\emailAdd{Wafaa\textunderscore Saleh@cis.asu.edu.eg}
\emailAdd{asmaa.shalaby@fsc.bu.edu.eg}

\abstract{ The transverse momentum distribution of charged particles formed in Au$–$Au collisions at Beam Energy Scan (BES) ($\sqrt{S_{NN}}= 7.7, 11.5, 19.6, 27, 39 \hspace{0.05cm} GeV$) is investigated. In addition, $ P_{T}$ spectra of $\Lambda$ particle at $\sqrt{S_{NN}}=62.4 \hspace{0.03cm} GeV$ was examined. Tsallis distribution  is used to extract the temperature, volume and the entropic index from the experimental results at mid-rapidity and zero chemical potential. We measure some particle ratios like $ \frac{k^{+}}{\pi^{+}} $ and $ \frac{\Lambda}{\pi^{-}} $ which are puzzling horn in the experiment and in the thermal model. We conclude that the horn vanished when we used Tsallis distribution, but this does not confirm a solution to the puzzle, which is primarily visible in the experimental results.} 


\begin{document}

\maketitle
\flushbottom

\section{Introduction}
Thermal models have been successful in describing particle yields at different beam energies 
\cite{cleymans2006comparison,andronic2006hadron,becattini2006energy} especially in heavy ion collisions, and many others
 \cite{tawfik2015particle,tawfik2016particle}, and under the effect of the magnetic field \cite{endrHodi2013qcd,bali2012qcd,tawfik2016qcd}. These models assume the formation of a system which is in thermal and chemical equilibrium in the hadronic phase and are characterized by a set of thermodynamic variables for the hadronic phase, most important among these are the chemical freeze-out temperature and baryon chemical potential

The transverse momentum spectrum of hadrons is considered as a consequence of the transverse flow of hot matter. It has been measured for various identified particle species in (Nucleus-Nucleus) AA or (proton-proton) pp, (proton-nucleus) pA collisions \cite{Yagi:2005yb}. In general, the single–particle spectra are well behaved  with the invariant cross section $\sigma$, in which the later is described by an exponential form in terms of the transverse mass, $m_{T}$ or $ m_{T}-m$ within  the transverse momentum $p_{T} < 2\hspace{0.04cm} GeV/c$ such that:

\begin{equation}
\frac{1}{2\hspace{0.02cm}\pi \hspace{0.03cm}m_{T}} \frac{d^{2}\sigma}{dm_{T} \hspace{0.03cm} dy} \approx exp\left( -m_{T}/T\right) 
\end{equation}
Where, y is the rapidity, T is the temperature and the transverse mass, $m_{T} = \sqrt{m^{2}+p_{T}^{2}}$ with the particle mass m. All units are taken as $\hbar = c = k_{B} = 1$.

The Tsallis statistics, in particular Tsallis distribution has get a significant role in the high energy physics. Especially, it fits the experimental data at $\sqrt{S_{NN}} = 200 \hspace{0.06cm}$ GeV \cite{abelev2007strange}, and at high energy $\sqrt{S_{NN}} = 0.9, 7  \hspace{0.06cm}$  TeV \cite{cleymans2011tsallis}, and at 0.9 \hspace{0.04cm} TeV \cite{aamodt2011production,cms2011strange}. Recently some of identified baryons and meesons are produced at the Beam Energy Scan (BES) ($\sqrt{S_{NN}} = $7.7, 11.5, 19.6, 27, 39 \hspace{0.06cm} GeV) \cite{adamczyk2017bulk}.

The production of pions, kaons and protons has been measured in pp and $Pb–Pb$ collisions at high $p_{T}$\cite{abelev2014production}, for $Au + Au$ collision \cite{abelev2007energy,adamczyk2017bulk}, and $p+Pb$\cite{abelev2014multiplicity}. The transverse momentum distribution of produced particles in AA collisions has been studied with Boltzmann-Gibbs distribution which takes the exponential form \cite{bashir2019particle,liu2013transverse,wang2019comparing}. In addition it has been studied using the non-extensive transverse momentum  distribution cite{rybczynski2020similarities,wei2016kinetic,parvan2017systematic}. In particular for $p_{T}$ spectra in pp collision \cite{marques2013nonextensivity,azmi2014transverse} in which the distribution for the whole charged particles were studied. In the present work we have studied the transverse momentum distribution for individual particles at BES energies \cite{adamczyk2017bulk}.
The validity domain of the standard statistical mechanics is examined and discussed many years ago \citep{di1942valoreE,shannon1948mathematical,tisza1966generalized,landsberg1984equilibrium} and \cite{maddox1993entropy,cohen2005boltzmann,cohen2002statistics}.

The motivation for examining of "new" statistics or a generalization of the standard statistics, in particular the non-extensive thermodynamics is coming from the contribution of "universality classes of systems" \cite{tsallis2009introduction}. Such as the complex systems, gravitational systems and long-range systems and many others in which they lose their extensive properties. For more details about the foundation of non-extensive statistics \cite{tsallis2009introduction} and also its applications see \cite{abe2001nonextensive}.

Eventually, in order to introduce Tsallis statistics, it is convenient to represent Boltzmann-Gibbs (BG) statistics first. Starting with BG distribution which is defined as the probability of finding the system in a certain state, for all microstates W the entropy is given as  \cite{tsallis2009introduction}

\begin{equation} \label{eq:entro1}
 S_{BG} = -k_{B} \sum_{i=1}^{W} p_{i} \hspace{0.05cm}ln p_{i}.           
\end{equation}

 Where $k_{B}$ is the Boltzmann's constant, and the total probability is the sum of probabilities for each particle i:
 
\begin{equation} \label{eq:prop}           
\sum_{i=1}^{W} p_{i} = 1.
\end{equation}

Equation \ref{eq:entro1} can be rewritten in the well-known entropy form \cite{boltzmann,gibbs}.

\begin{equation} \label{eq:entro2}
              S_{BG} = k_{B} \hspace{0.05cm}ln W.
\end{equation}

The probability can be reformulated in terms of the particle energy, $E_{i}$ , temperature, T, and the partition function $\textit{ Z}_{BG}$,
\begin{equation} \label{eq:prop1}
  p_{i}= \frac{e^{-\beta E_{i}}}{ \textit{Z}_{BG}  }, \quad \hspace{0.05cm}  \beta \equiv \frac{1}{k_{B}\hspace{0.03cm} T}
\end{equation}
  Where 
\begin{equation} \label{eq:parti}
    \textit{ Z}_{BG}\equiv \sum_{i=1}^{W} e^{-\beta E_{i}}.
\end{equation}
where $ E_{i} = \sqrt{p^2+m_{i}^{2}}$, is the energy of the $i^{th}$ particle, and $m_{i}$ the particle mass. Now we represent the most important characteristic features of BG systems is the additive property. Simply, it can be defined as follows : for a system \textbf{A} composed of two subsystems $\textbf{A}_{1}$ and $\textbf{A}_{2}$, the total entropy can be written as,

\begin{equation} 
                S _{BG}(A)= S _{BG}(A_{1}) + S _{BG}(A_{2}).
\end{equation}

On the other hand the Tsallis statistics leads to non-additive property,

\begin{equation} \label{eq:entro4}
 S_{q}(A_{1}+A_{2})= S_{q}(A_{1})+ S_{q}(A_{2})+ (1-q)S_{q}(A_{1}) S_{q}(A_{2}).
\end{equation}
Where ,q, is the non-additive parameter, or called the entropic index, in which BG is recovered as $(q\rightarrow 1)$. Finally, the entropy is Eq.\ref{eq:entro1} is generalized in non-extensive form to be as follows \cite{1988JSP....52..479T}.

\begin{equation} \label {eq:entro5}
S_{q}\equiv k_{B} \frac{\left[1- \sum_{i=1}^{W} p_{i}^{q} \right] }{\left( q-1\right) }.
\end{equation}

\begin{equation} \label{eq:pati1}
p_{i}^{q} = \frac{\left[ 1-\beta\hspace{0.03cm}(q-1)\hspace{0.03cm}E_{i}\right] ^\frac{1}{q-1}}{\textit{Z}_{q}}   \quad \hspace{0.1cm},
\textit{Z}_{q} = \sum_{i=1}^{W}\left[ 1-\beta\hspace{0.03cm}(q-1)\hspace{0.03cm}E_{i}\right] ^\frac{1}{(q-1)}.
\end{equation}
Therefore, the parameter q would determine the non-additivity "degree" of the system, that BG can be restored as q tends to unity.
  
\section{Non-Extensive Distribution} \label{sec:Model}
In this section we introduce a brief discussion on Tsallis statistics in particular the number of particles and the transverse momentum distribution. Starting with the number of particles N, and number density n, \cite{cleymans2012relativistic} given by the following expressions:

\begin{eqnarray}\label{eq:num}
n_{i}=\frac{N_{i}}{V} = g_{i} \hspace{0.03cm} V \int \frac{d^{3}p}{\left( 2\hspace{0.02cm}\pi \right)^{3}} \hspace{0.04cm} f^{q} 
\end{eqnarray}
The total number density $ n = \sum_{i} n_{i}$, and $ N = \sum N_{i}$, where the Tsallis function $f^{q}$ is defined as \cite{1988JSP....52..479T,tsallis1998role,azmi2014transverse}. 

\begin{eqnarray}
f_{i}^{q} = \left[ 1+(q-1)\frac{E_{i}-\mu_{i}}{T}\right]^{\frac{-1}{q-1}}
\end{eqnarray}

Where $ g_{i}, T, V$ and $\mu_{i} $ are degeneracy factor, temperature, volume and chemical potential respectively. The latter is defined as $\mu_{i} = \mu_{b} \hspace{0.03cm} B_{i}+\mu_{s} \hspace{0.03cm}S_{i}+ \mu_{q} \hspace{0.03cm} Q_{i}$ for any particle i with $B_{i},S_{i}, Q_{i}$ are the baryon, strangeness, and charge quantum numbers, and the corresponding chemical potentials are $\mu_{b}, \mu_{s}, \mu_{q}$ respectively. In the present work we have performed all calculations at zero chemical potential.

 Now we turn to represent the transverse momentum distribution based on Tsallis distribution \cite{conroy2010thermodynamic, cleymans2012relativistic}.
 \begin{equation} \label {eq:dist2}
 E \hspace{0.03cm} \frac{d^{3}N}{dp^{3}} =  \frac{g \hspace{0.02cm}V \hspace{0.02cm} E}{(2\pi)^{3}} \left[ 1+(q-1)\frac{E-\mu}{T}\right] ^{\frac{-q}{q-1}}
\end{equation}
 Eq. (\ref{eq:dist2}) can be re-written in terms of the rapidity $ y $, and transverse mass $ m_{T} =\sqrt{m^{2}+p_{T}^{2}} $ variables. Given $ ( E = m_{T} cosh y) $  \cite{azmi2014transverse} :

\begin{equation} \label{eq:ptdist}
\frac{d^2 N}{dp_T dy}=gV\frac{p_T m_T \cosh y}{(2\pi)^2}\left[1 + (q-1)\frac{m_T{ cosh}y - \mu}{T}\right]^{-q/(q-1)}
\end{equation}
At mid-rapidity, y = 0, and for zero chemical potential, eq.(\ref{eq:ptdist}) takes the following form:
\begin{eqnarray}\label{eq:fitrel}
\left.\frac{d^{2}N}{2\hspace{0.03cm}\pi \hspace{0.03cm}p_{T}\hspace{0.03cm}dp_T~dy}\right|_{y=0} = gV\frac{p_Tm_T}{(2\pi)^3}\left[1+(q-1)\frac{m_T}{T}\right]^{-q/(q-1)}
\end{eqnarray}
We have utilized Eq. \ref{eq:fitrel} to fit the experimental data \cite{adamczyk2017bulk} by performing 
$ \chi^{2} $ fit.

\section{Results and Discussion}
\begin{figure}[H] 
     \centering 
      \setlength\abovecaptionskip{-0.05\baselineskip} 
 \subfigure[ ]{\label{fig:kp7_7}\includegraphics[height=5cm, width=7cm]{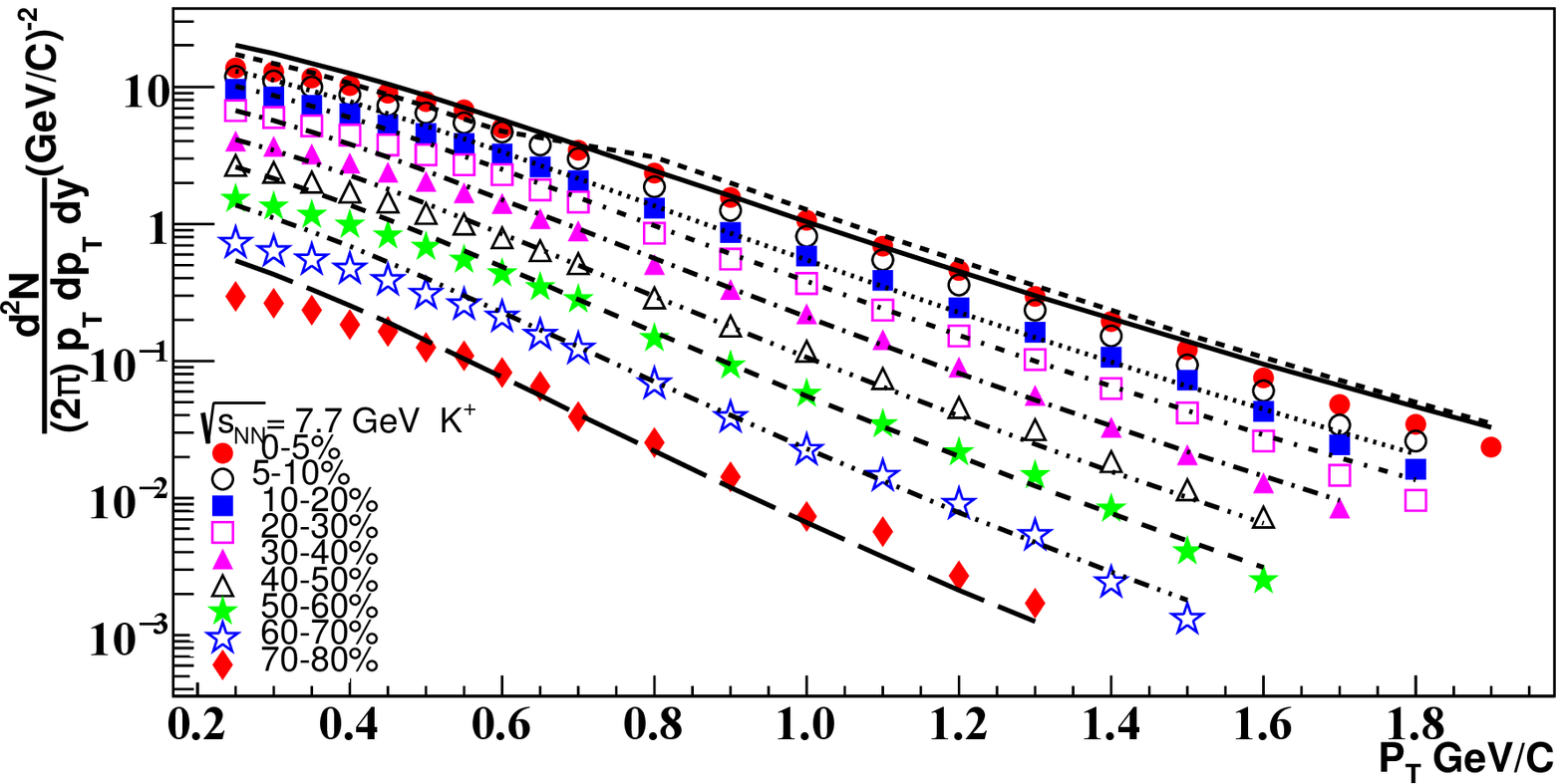}}\qquad
  \subfigure[ ]{\label{fig:km7_7}\includegraphics[height=5cm, width=7cm]{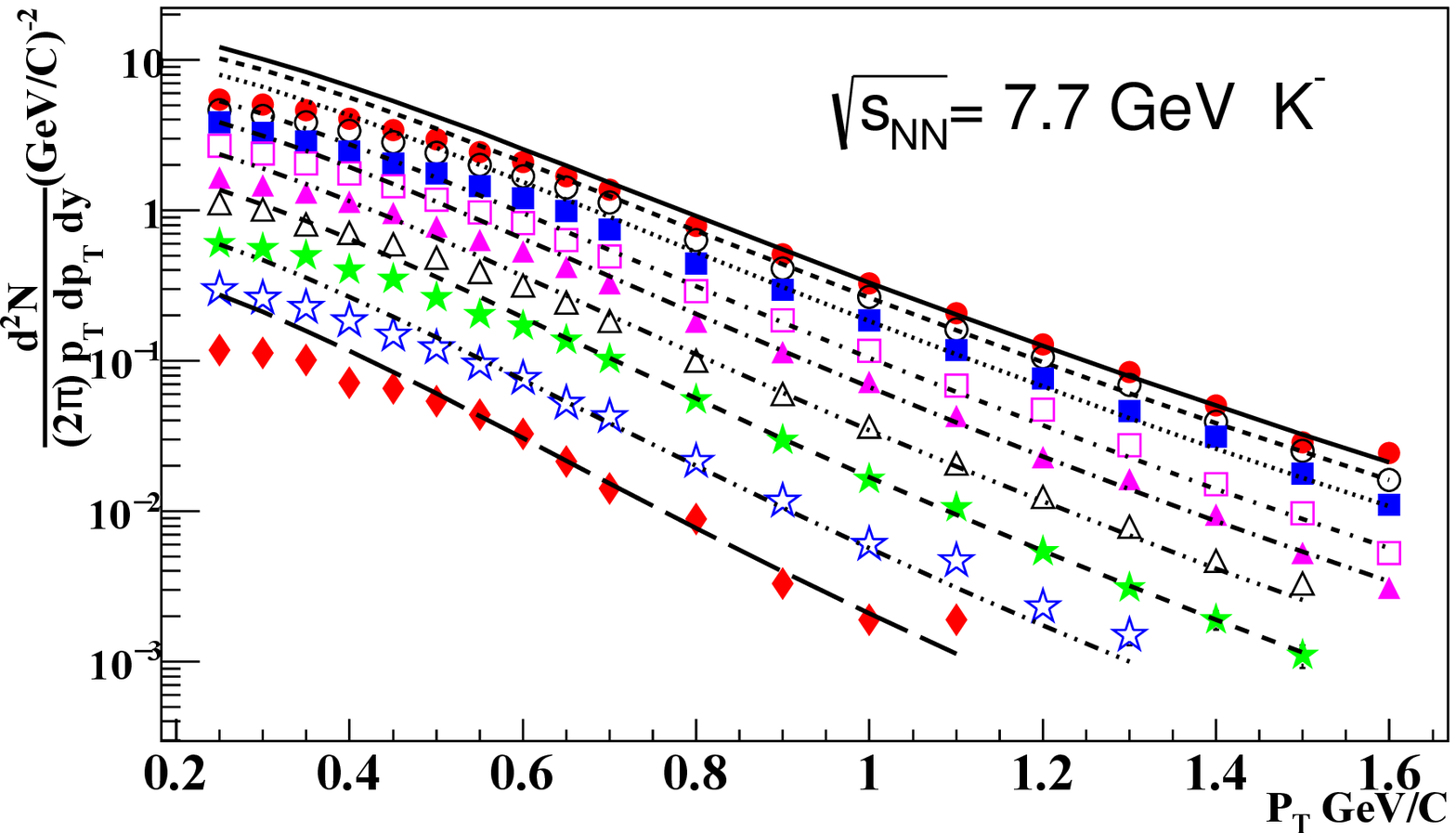}}\qquad  
   \subfigure[ ]{\label{fig:pip7_7}\includegraphics[height=5cm, width=7cm]{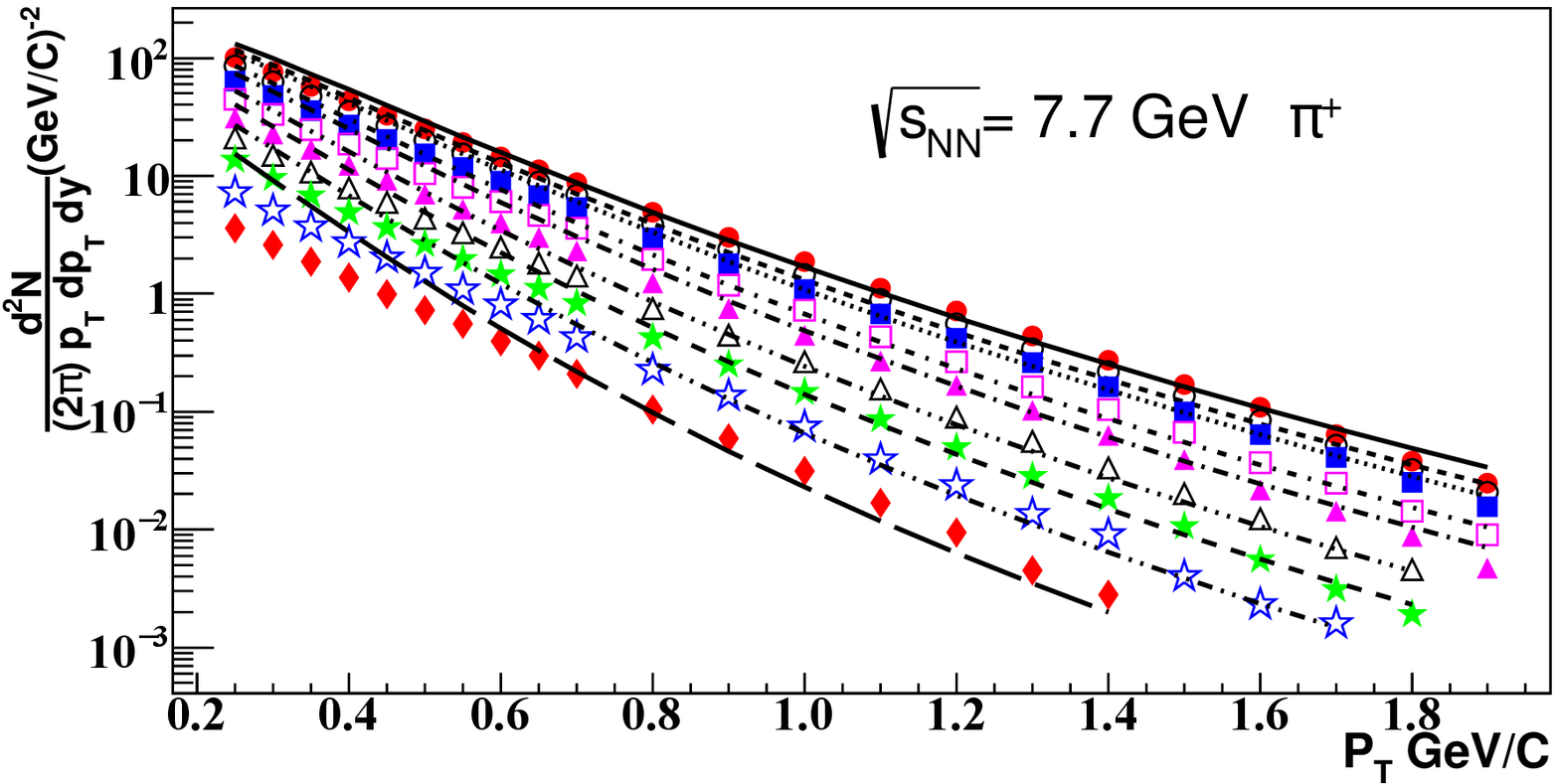}}\qquad  
 \subfigure[ ]{\label{fig:pim7_7}\includegraphics[height=5cm, width=7cm]{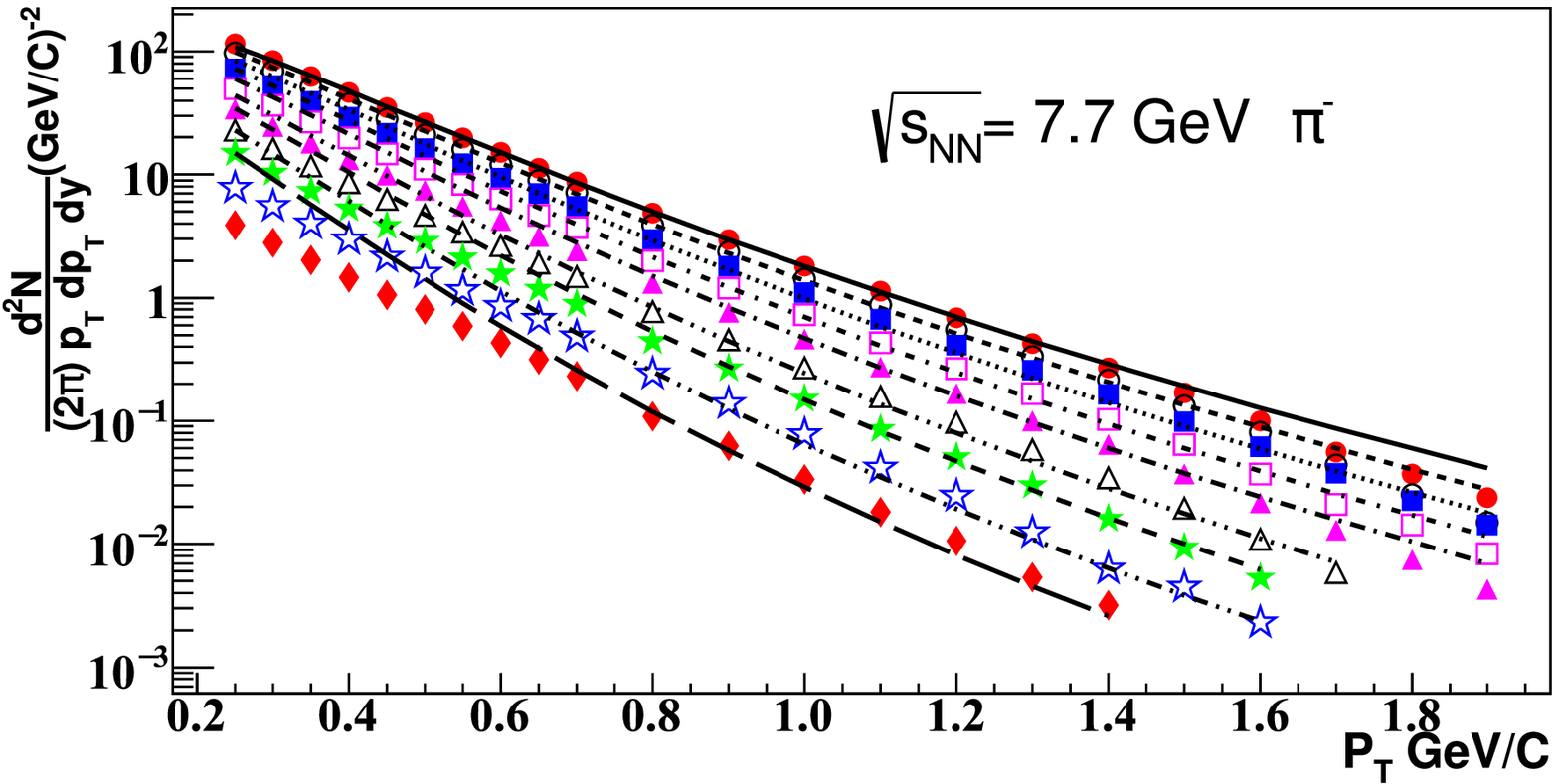}}\qquad 
 \subfigure[ ]{\label{fig:pbar7_7}\includegraphics[height=5cm, width=7cm]{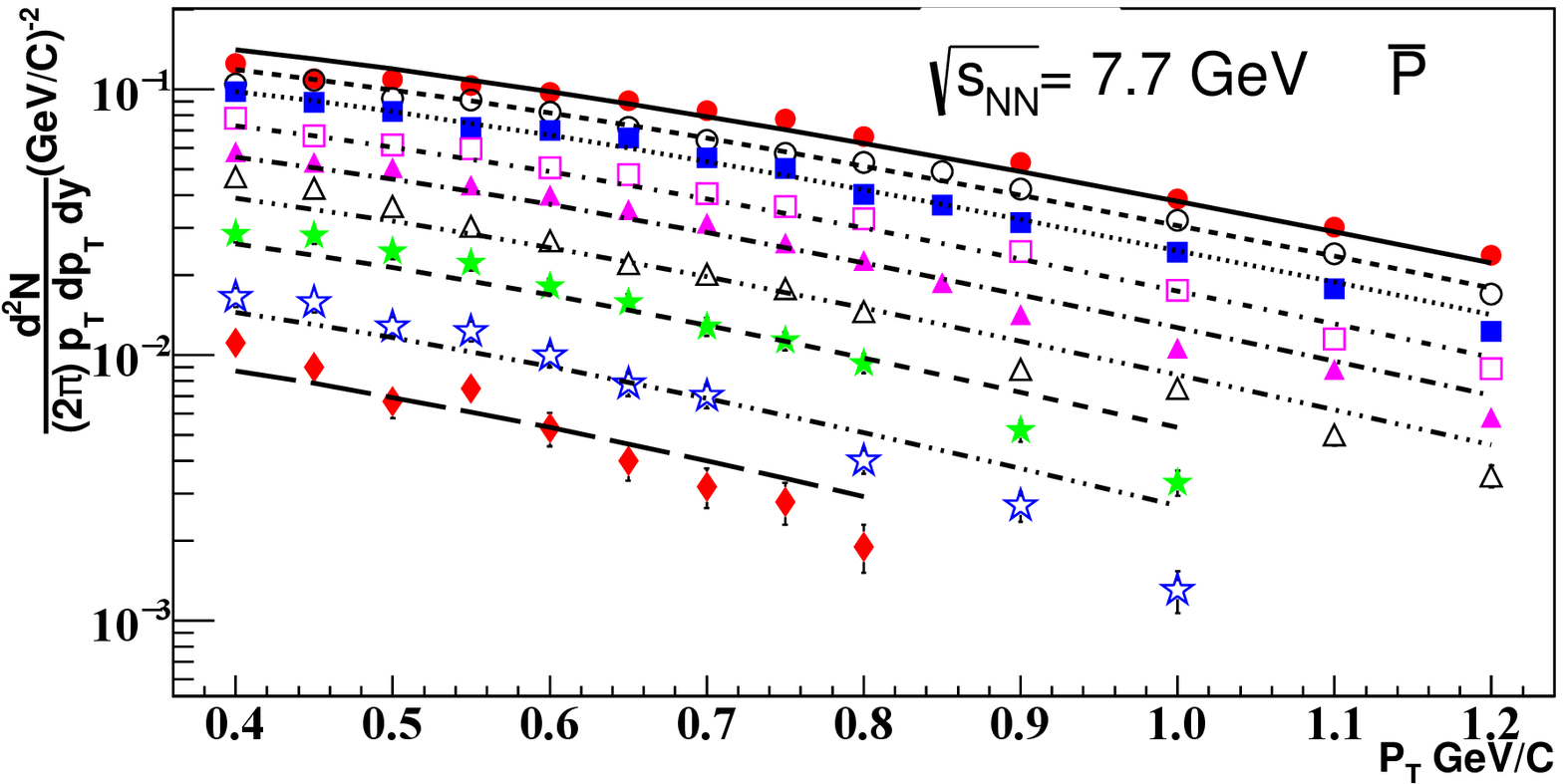}}\qquad  
 \subfigure[ ]{\label{fig:proton7_7}\includegraphics[height=5cm, width=7cm]{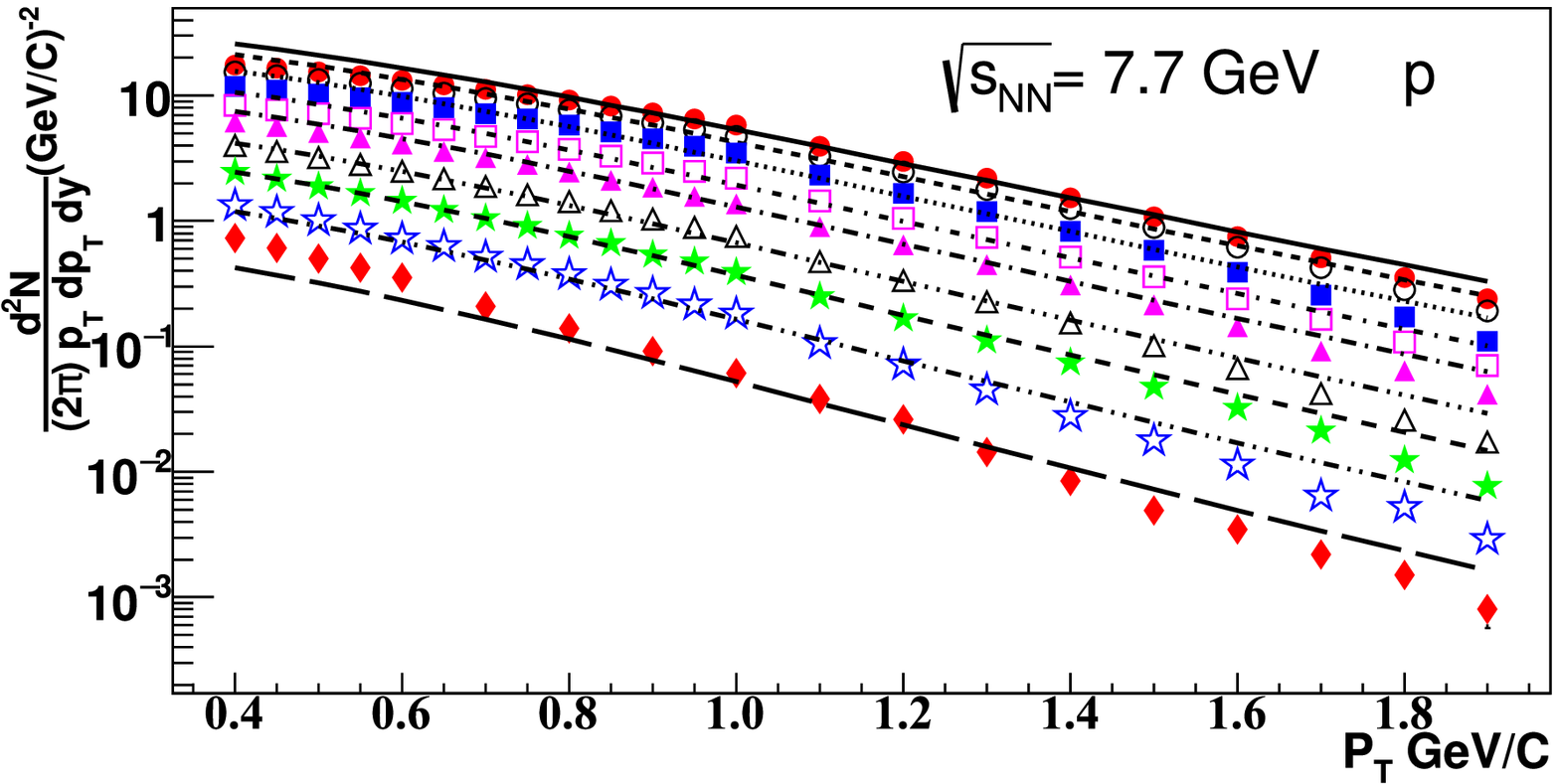}}\qquad 
 \caption[figtopcap]{The Colored-online solid and different dashed lines represent the transverse momentum distribution $ d^{2}N/2\pi P_{T} dP_{T} dy $ versus the transverse momentum $p_{T}$ for $k^{+}$, $k^{-}$, $\overline{p}$, $ p $ , $ \pi^{+} $ and $ \pi^{-} $ in Au+Au collisions at $\sqrt{s_{NN}} = 7.7 \hspace{0.05cm}GeV$ for different centralities. The Colored-online symbols are the corresponding experimental data \cite{adamczyk2017bulk}.}
  \label{fig:7_7kaon}
\end{figure}

\begin{figure}[H] 
     \centering 
      \setlength\abovecaptionskip{-0.05\baselineskip} 
 \subfigure[ ]{\label{fig:kp39}\includegraphics[height=5cm, width=7cm]{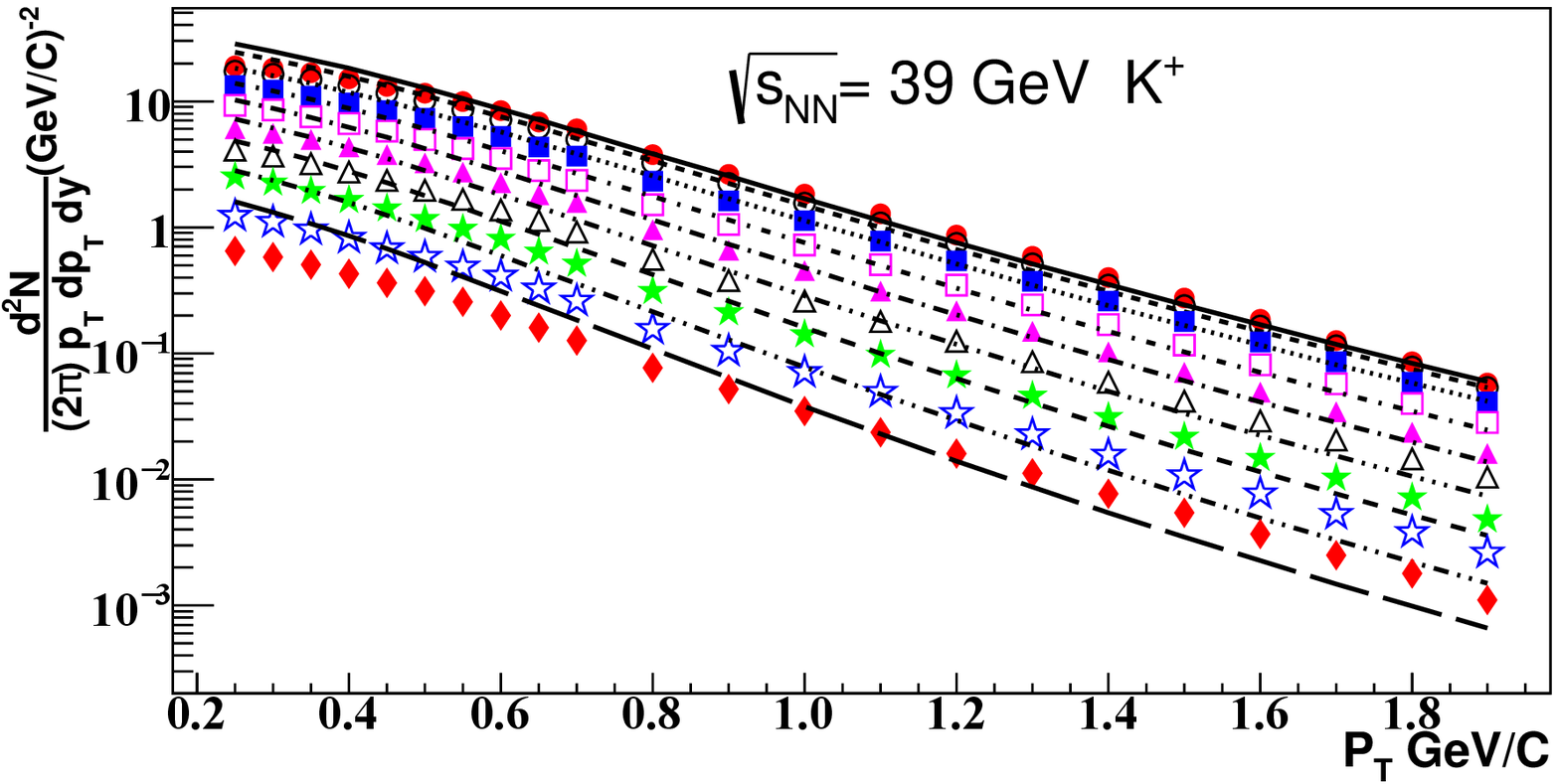}}\qquad
  \subfigure[ ]{\label{fig:km39}\includegraphics[height=5cm, width=7cm]{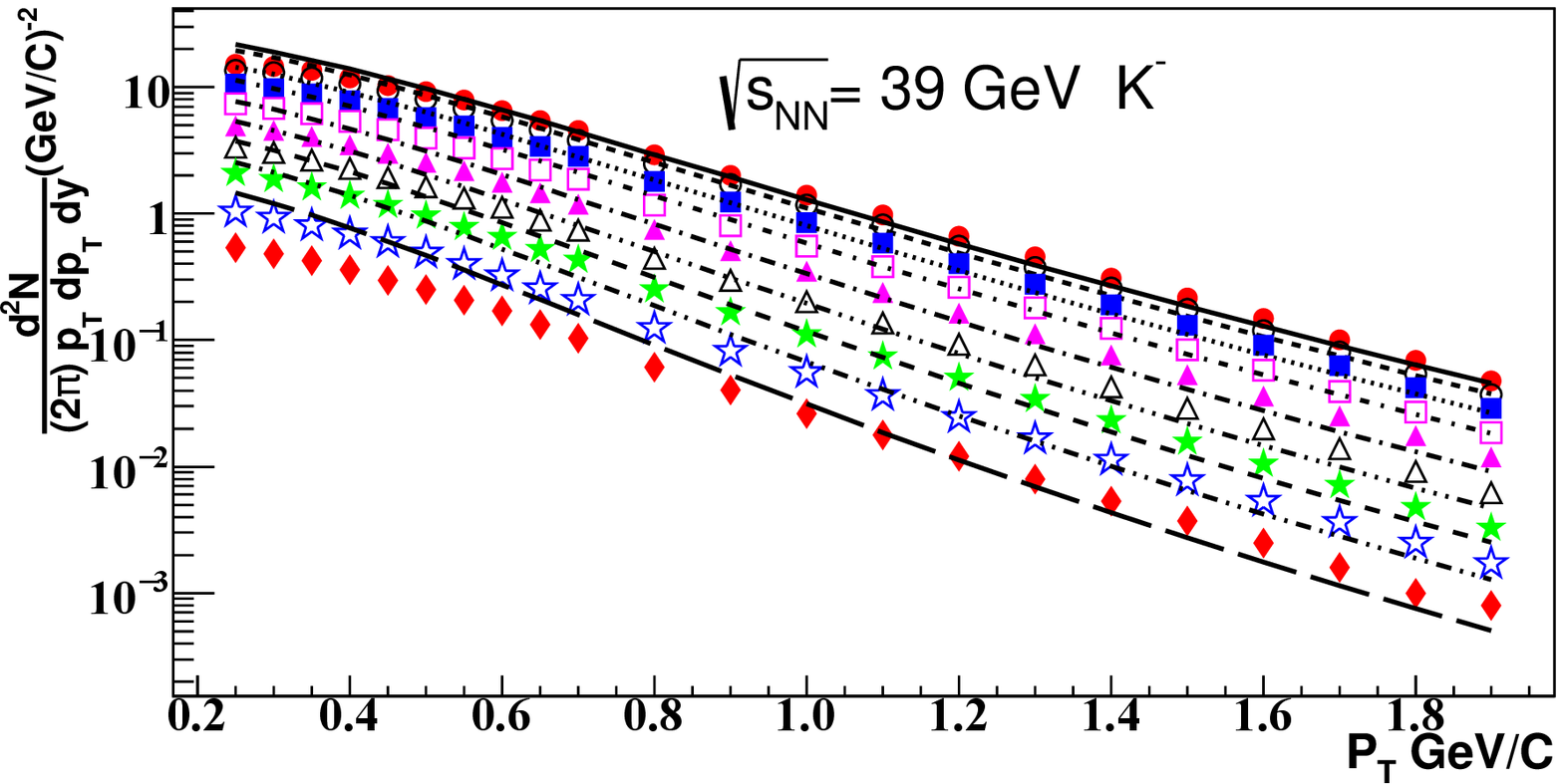}}\qquad  
 \subfigure[ ]{\label{fig:pip39}\includegraphics[height=5cm, width=7cm]{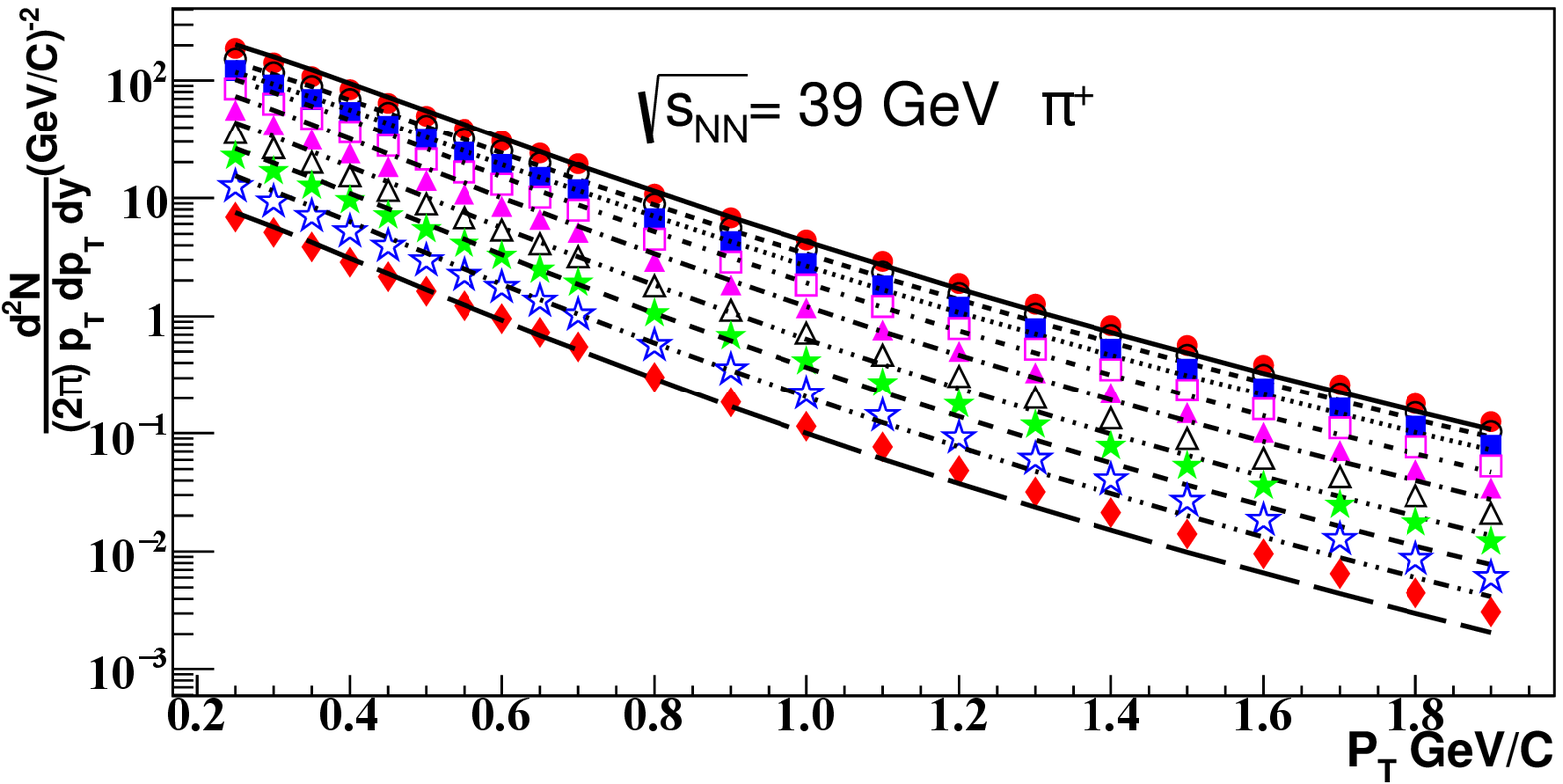}}\qquad  
 \subfigure[ ]{\label{fig:pim39}\includegraphics[height=5cm, width=7cm]{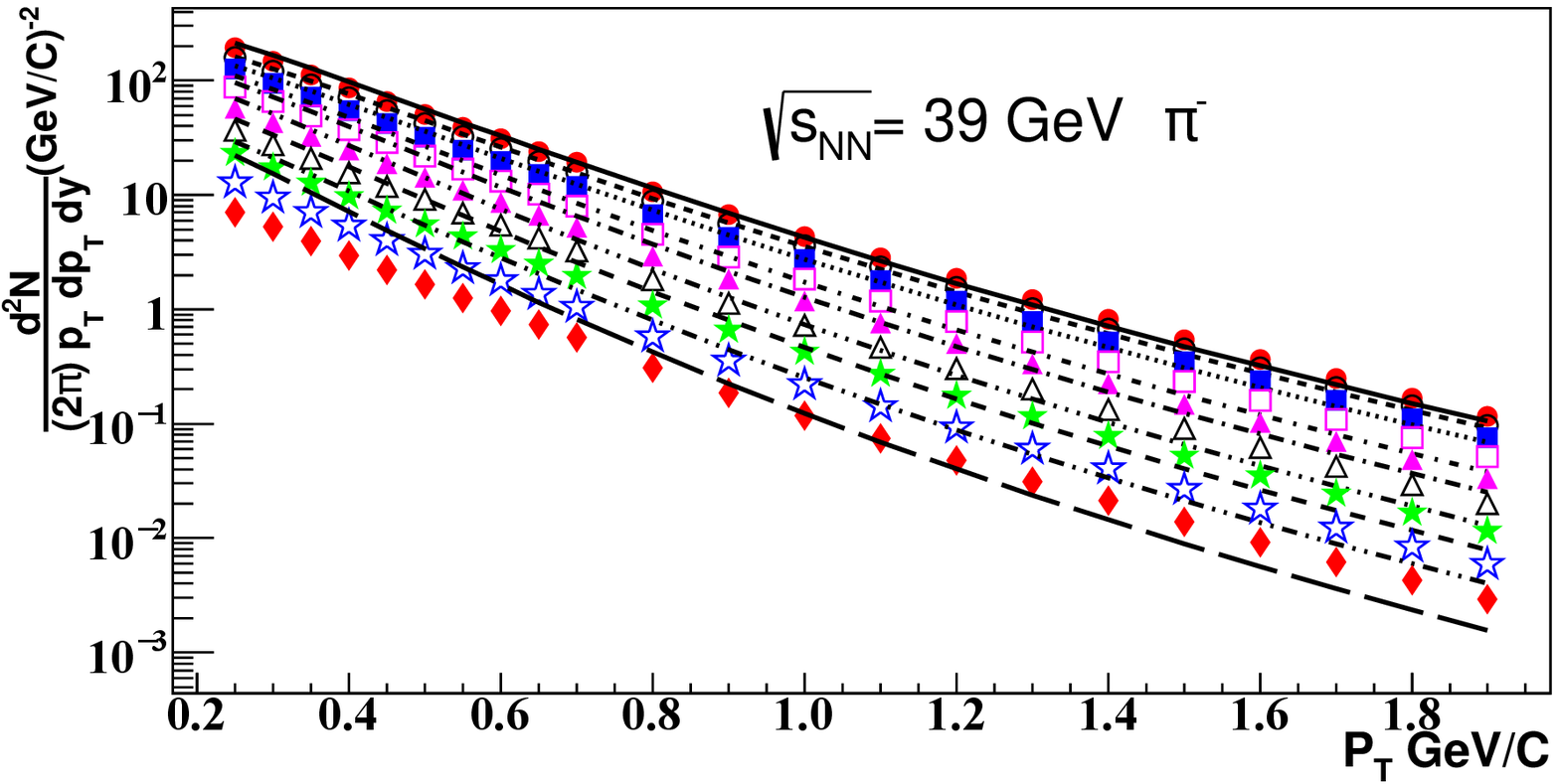}}\qquad 
   \subfigure[ ]{\label{fig:pbar39}\includegraphics[height=5cm, width=7cm]{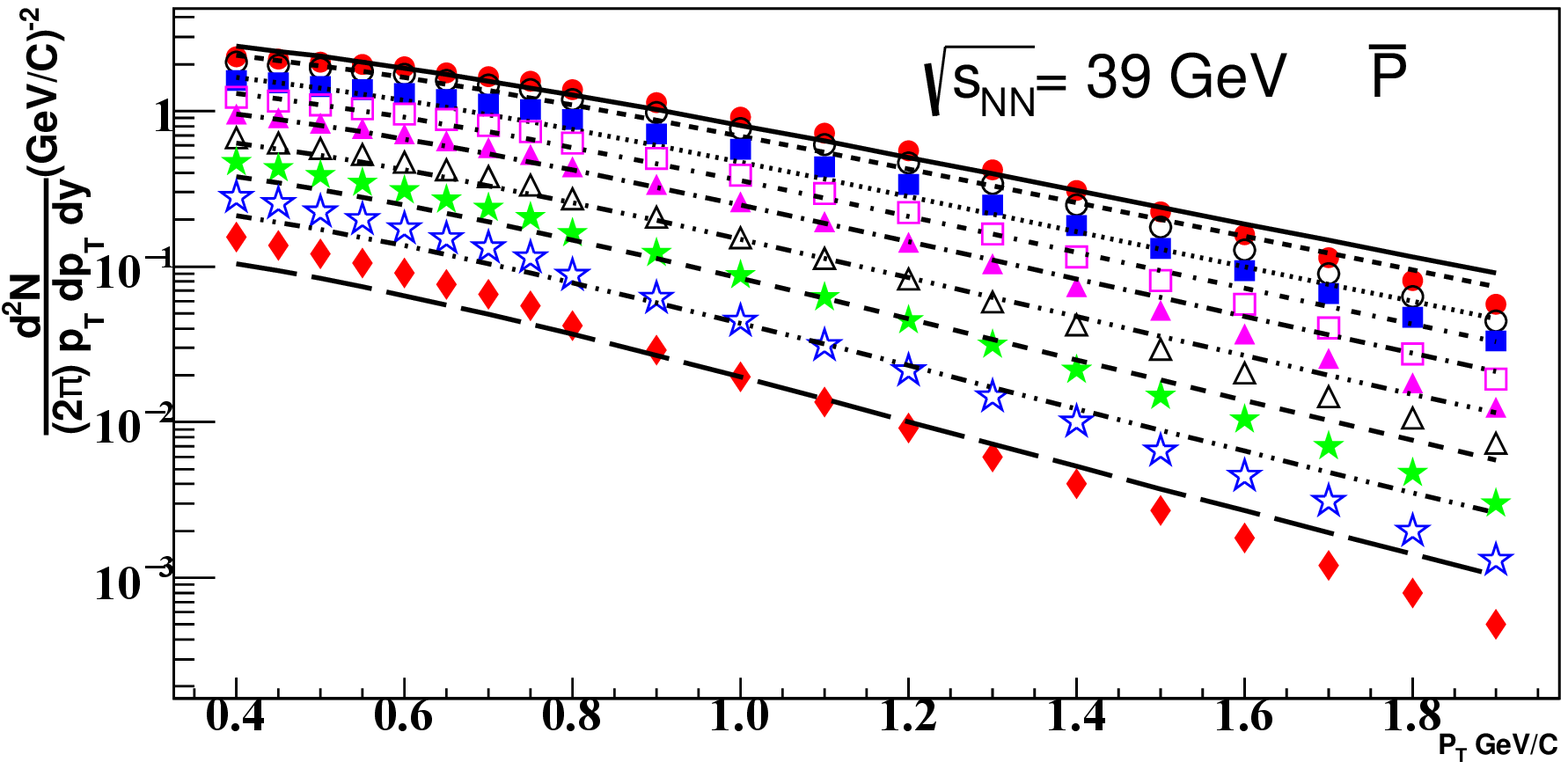}}\qquad  
 \subfigure[ ]{\label{fig:proton39}\includegraphics[height=5cm, width=7cm]{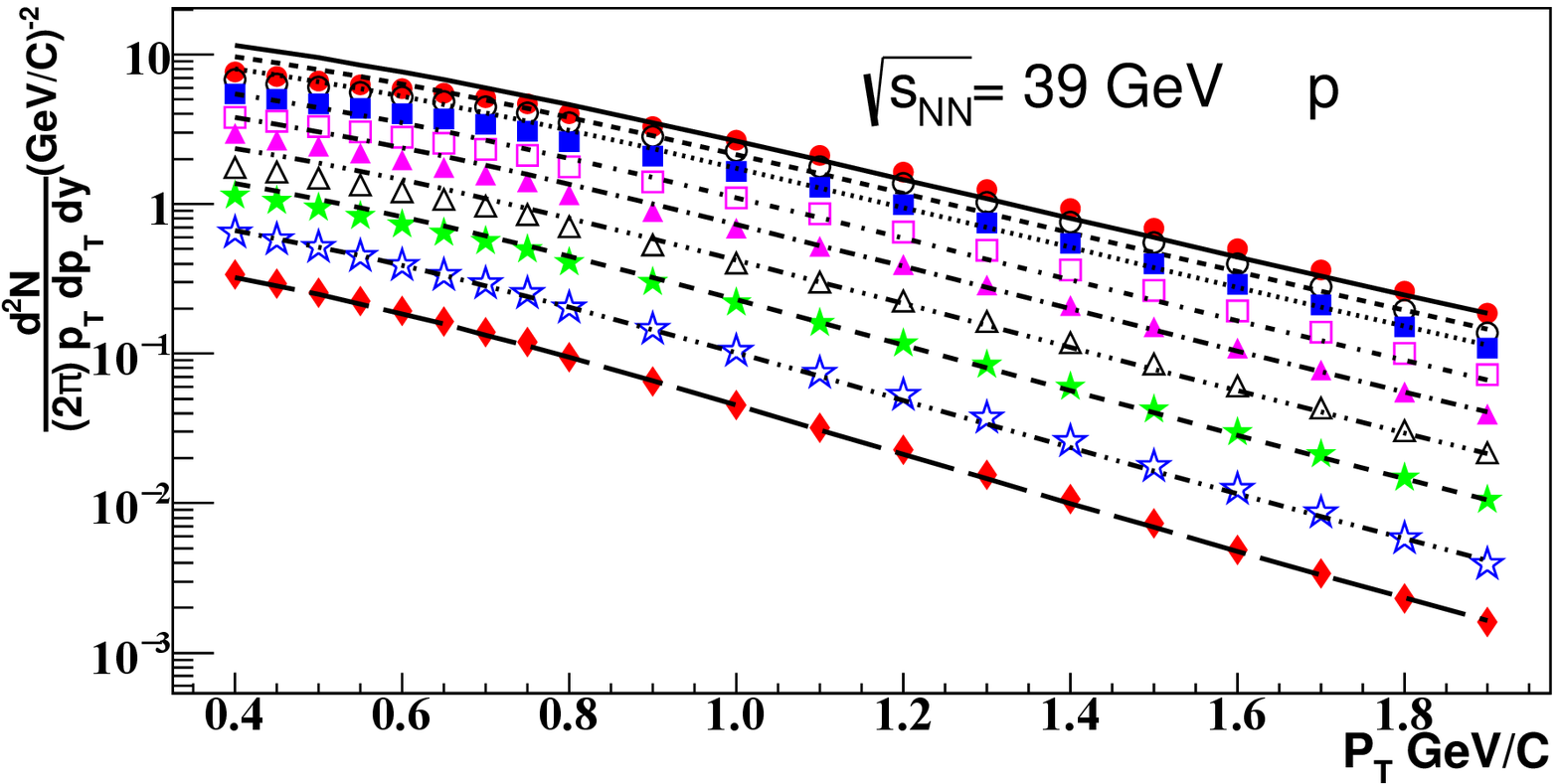}}\qquad 
 \caption[figtopcap]{The same as Fig (\ref{fig:7_7kaon}) at $\sqrt{s_{NN}} = 39 \hspace{0.05cm}GeV$}
  \label{fig:39kaon}
\end{figure}

 In this section we represent the transverse momentum distributions of identified particles 
 such as ($k^{\pm}, \pi^{\pm}, p, \overline{p}$). These particles are produced in Relativistic Heavy-Ion Collisions(RHIC)from the Beam Energy Scan Program (BES) at various center of mass energies,
 
 ($\sqrt{S_{NN}}= 7.7, 11.5, 19.6, 27, 39 \hspace{0.05cm} GeV$) 
 \cite{adamczyk2017bulk}, and for $\Lambda$ at $\sqrt{S_{NN}}=62.4 \hspace{0.03cm} GeV$ \cite{aggarwal2011strange}. The experimental results were fitted at rapidity $y = 0$ and zero chemical potential, using Tsallis distributions at fixed entropic index ($q = 1.09$).  Figure(\ref{fig:7_7kaon}), illustrates the Tsallis transverse momentum spectra for the identified particles ($k^{\pm}, \pi^{\pm}, p, \overline{p}$) for different centrality classes ($ 0-5 \%, 5-10 \%, 10-20 \%, 20-30 \%, 30-40 \%, 40-50 \%, 50-60 \%, 60-70 \%, 70-80 \%  $). The theoretical results are represented in different types of lines for the centrality classes, respectively and the corresponding colored symbols represent the experimental one. It is noticeable that, at $ 0.5 \lesssim p_{T}\lesssim 1.5 \hspace{0.05cm} GeV/c$ the distribution for the mentioned particles fits well with the experimental data. But slightly deviates at larger $p_{T}$.
The following figures (\ref{fig:11_5kaon}, \ref{fig:19_6kaon}, \ref{fig:27kaon}, \ref{fig:39kaon}) represent the same relations at other energies $\sqrt{S_{NN}}= 11.5, 19.6, 27, 39\hspace{0.05cm} GeV$. The reasonable fitting is clearly appeared by increasing the collision energy for all particle species except the deviation occurs at centrality ($ 70-80 \% $).
\newpage

\begin{figure}[H] 
     \centering 
      \setlength\abovecaptionskip{-0.05\baselineskip} 
 \subfigure[ ]{\label{fig:kp11_5}\includegraphics[height=5cm, width=7cm]{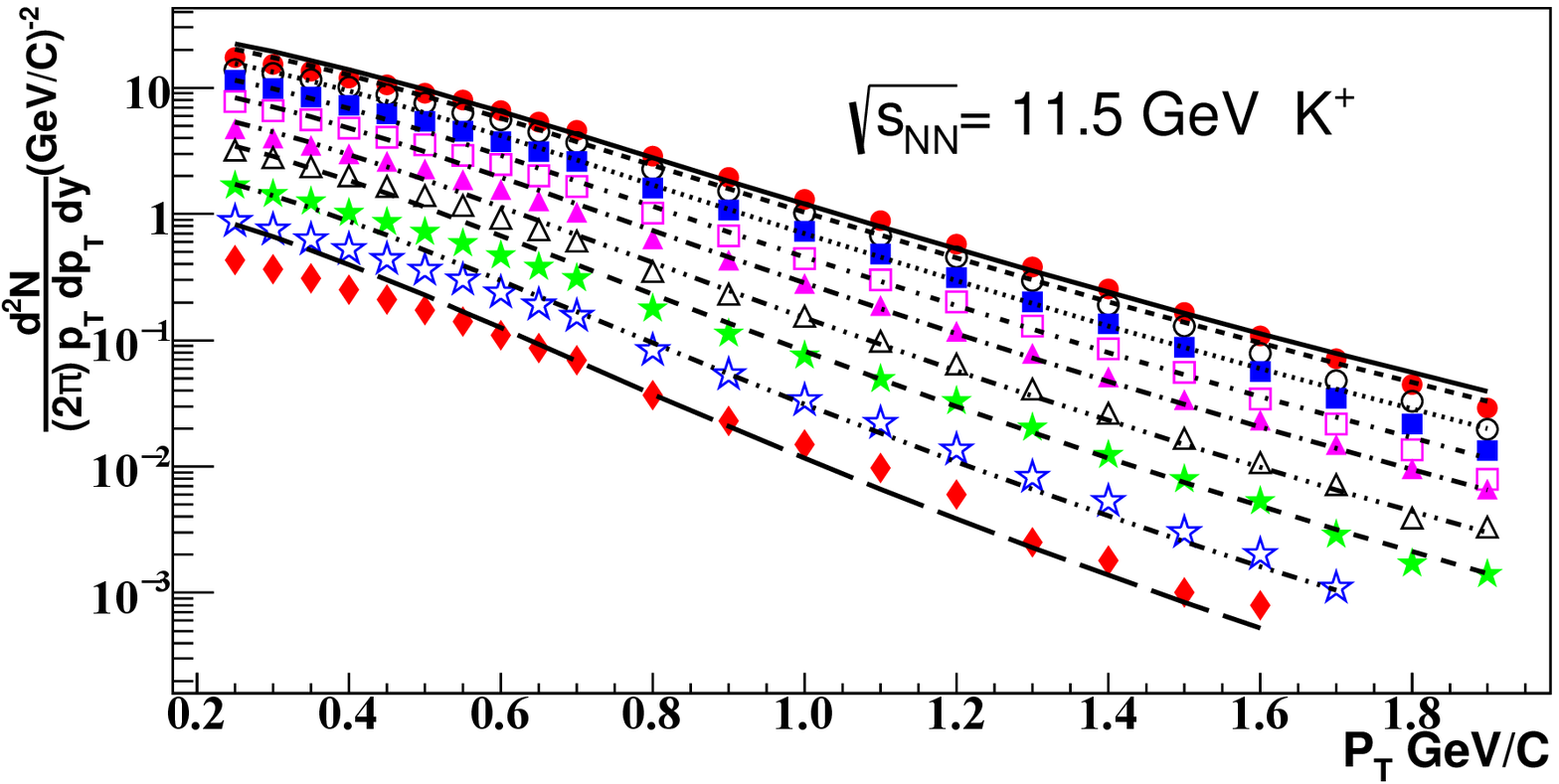}}\qquad
  \subfigure[ ]{\label{fig:km11_5}\includegraphics[height=5cm, width=7cm]{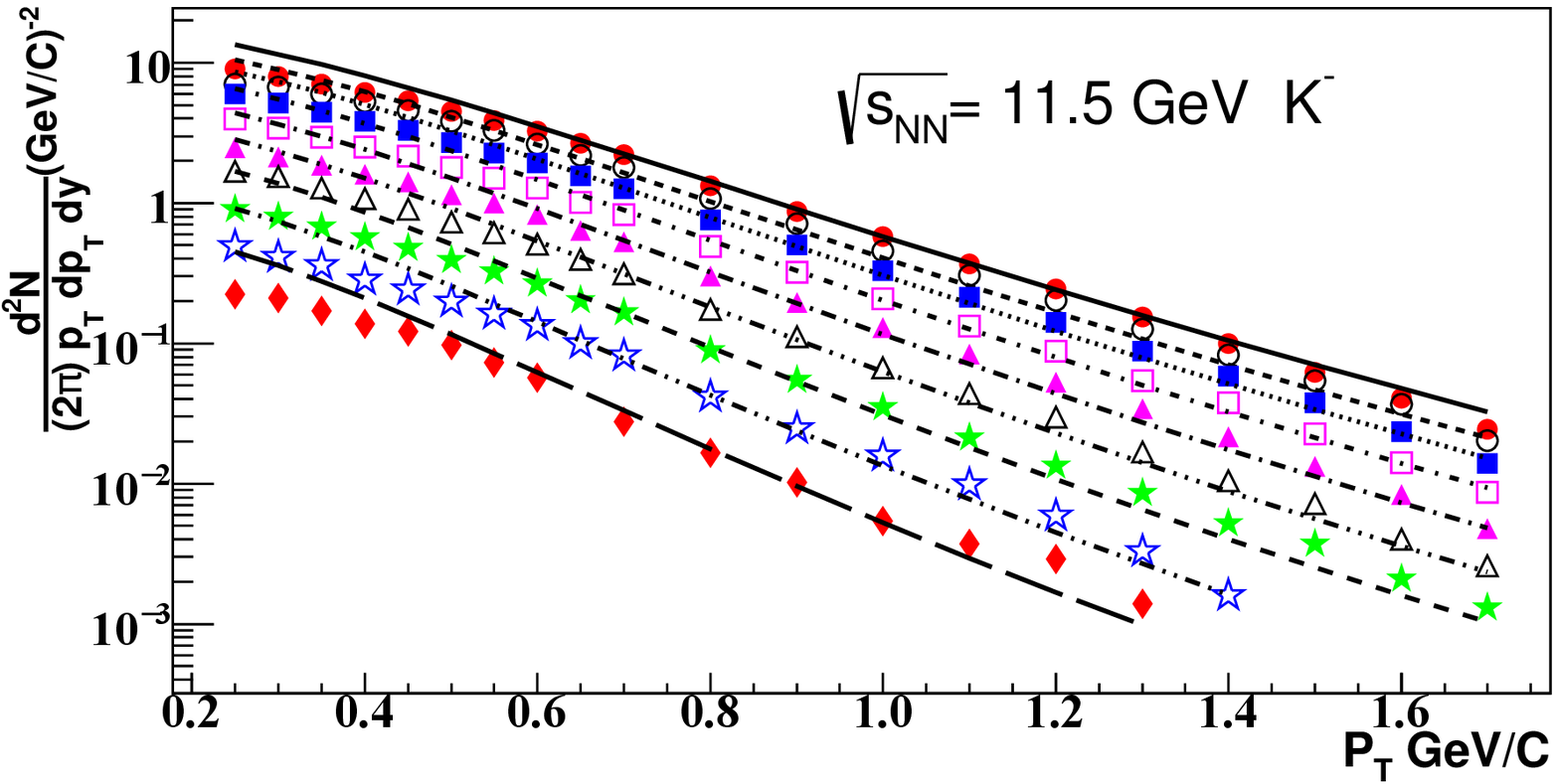}}\qquad  
   \subfigure[ ]{\label{fig:pip11_5}\includegraphics[height=5cm, width=7cm]{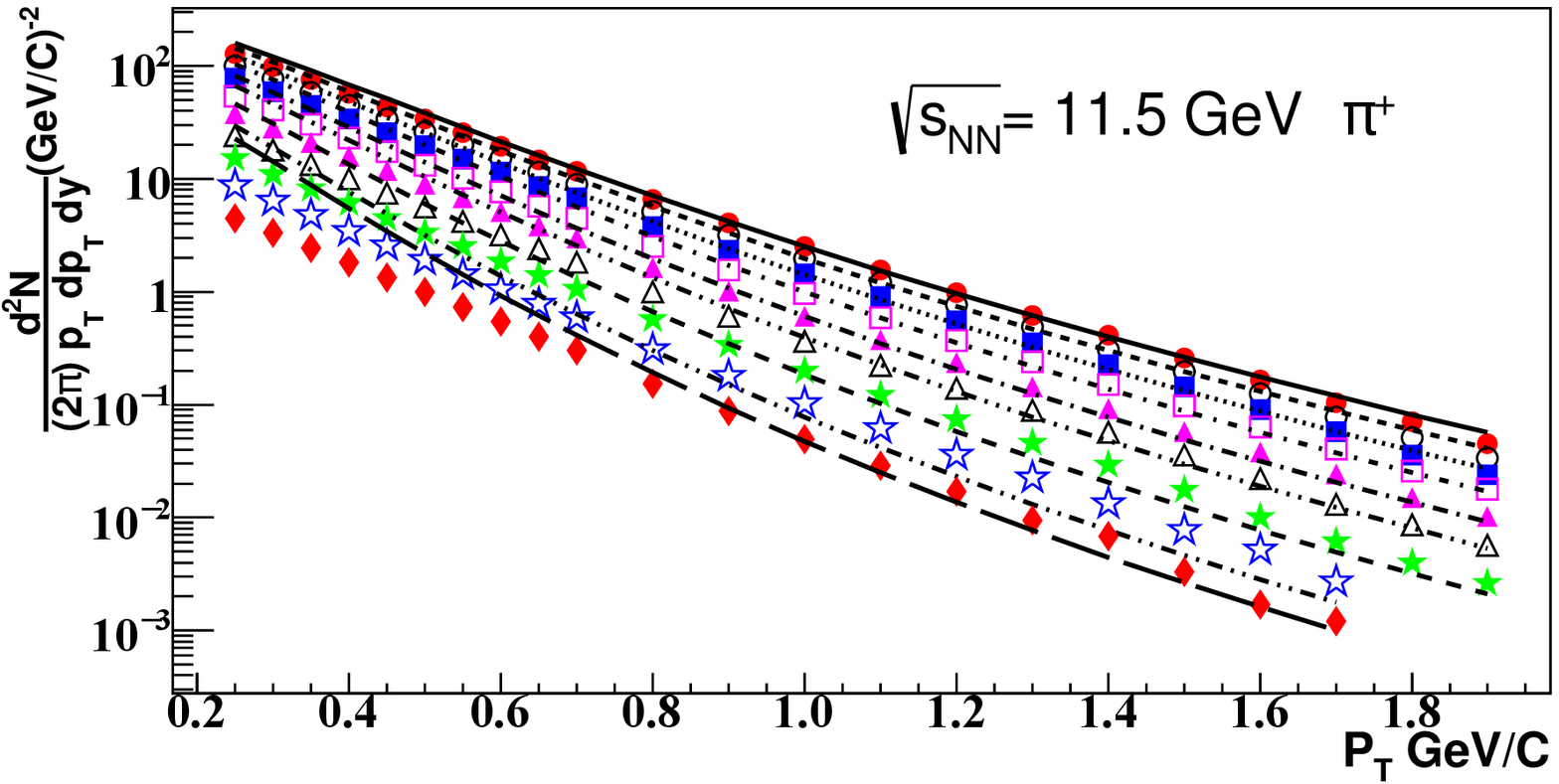}}\qquad  
 \subfigure[ ]{\label{fig:pim11_5}\includegraphics[height=5cm, width=7cm]{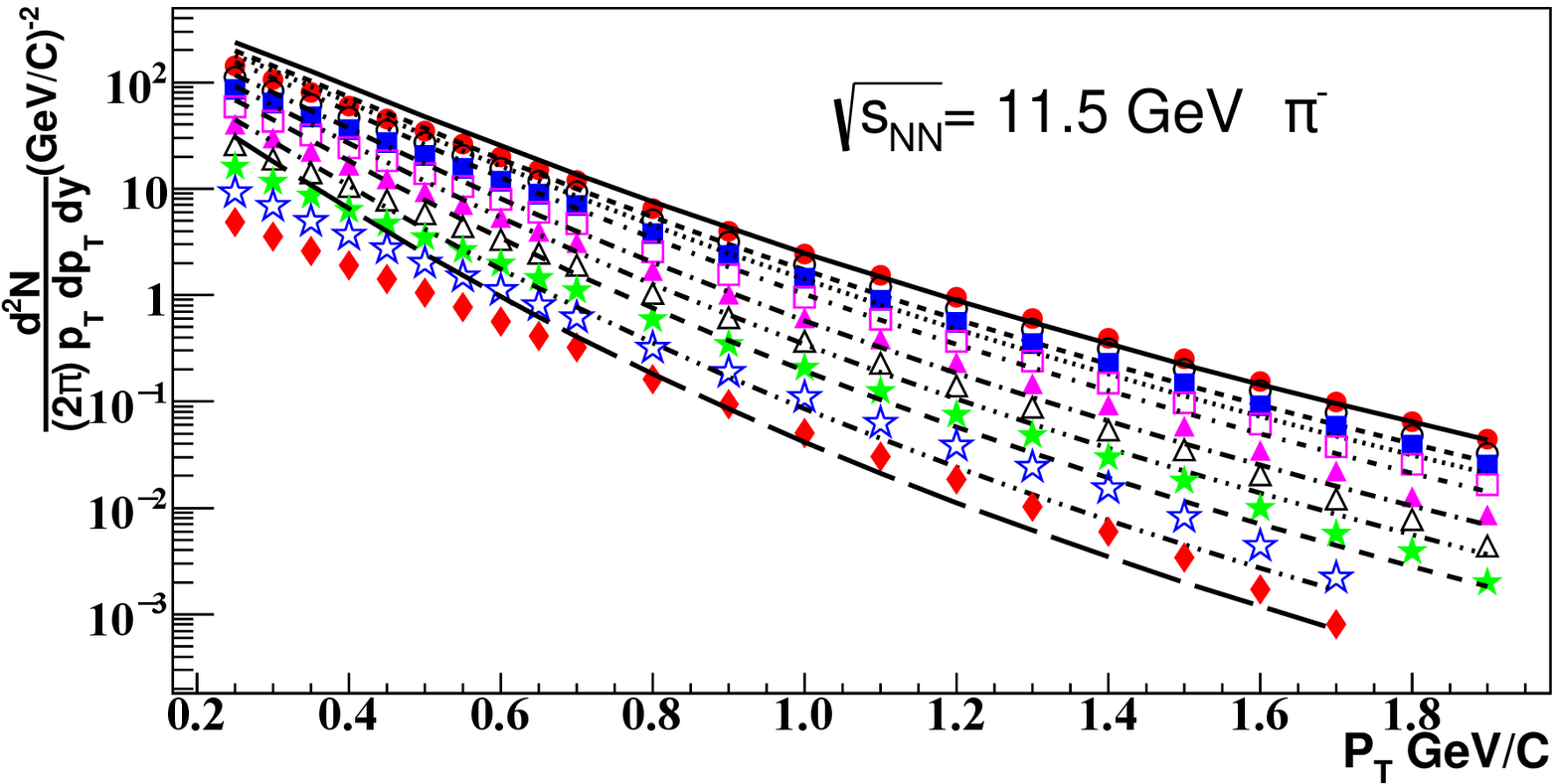}}\qquad 
 \subfigure[ ]{\label{fig:pbar11_5}\includegraphics[height=5cm, width=7cm]{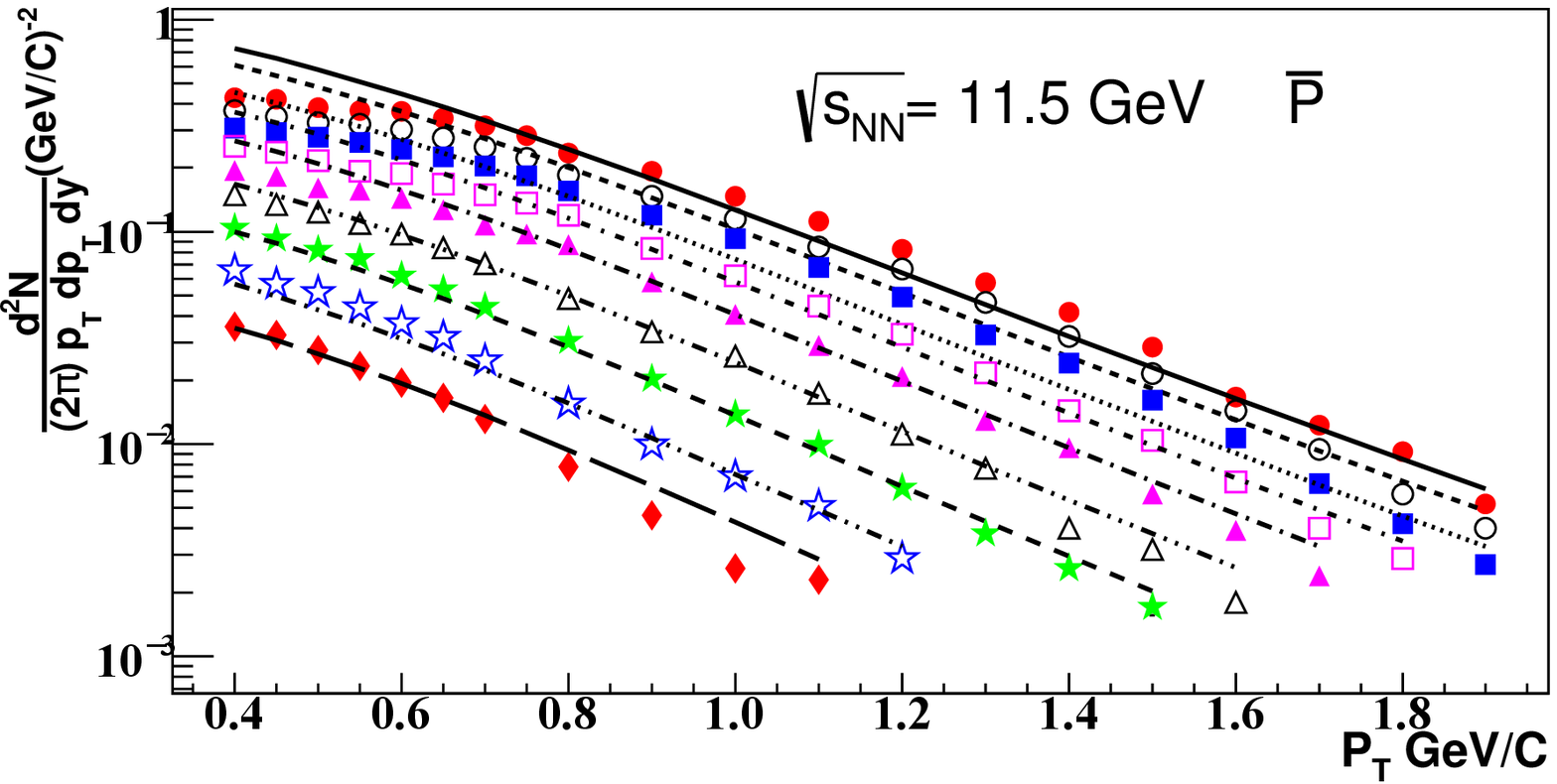}}\qquad  
 \subfigure[ ]{\label{fig:proton11_5}\includegraphics[height=5cm, width=7cm]{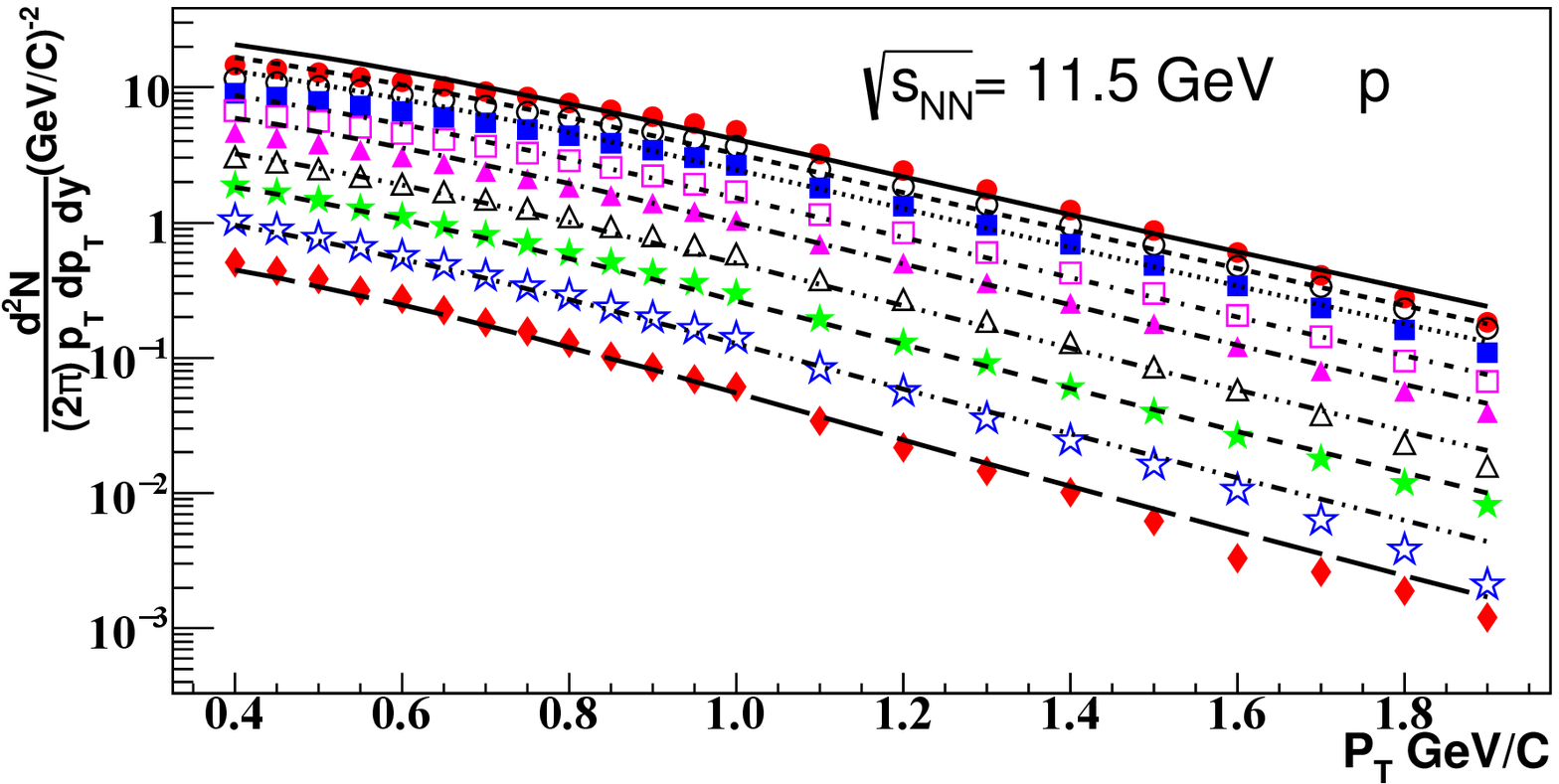}}\qquad 
 \caption[figtopcap]{ The same as Fig (\ref{fig:7_7kaon}) at $\sqrt{s_{NN}} = 11.5 \hspace{0.05cm}GeV$.}
  \label{fig:11_5kaon}
\end{figure}

\begin{figure}[H] 
     \centering 
      \setlength\abovecaptionskip{-0.05\baselineskip} 
 \subfigure[ ]{\label{fig:kp19_6}\includegraphics[height=5cm, width=7cm]{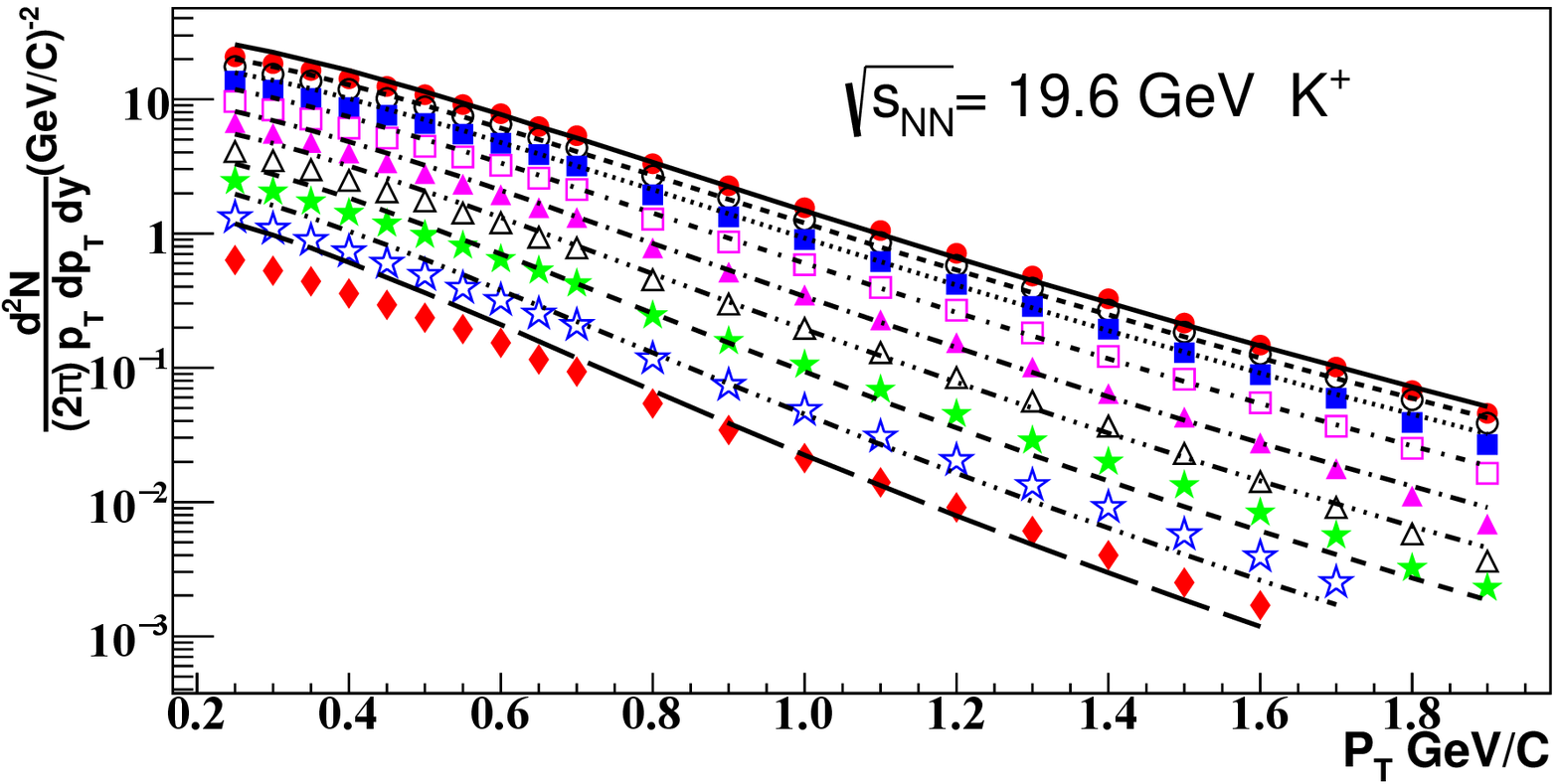}}\qquad
  \subfigure[ ]{\label{fig:km19_6}\includegraphics[height=5cm, width=7cm]{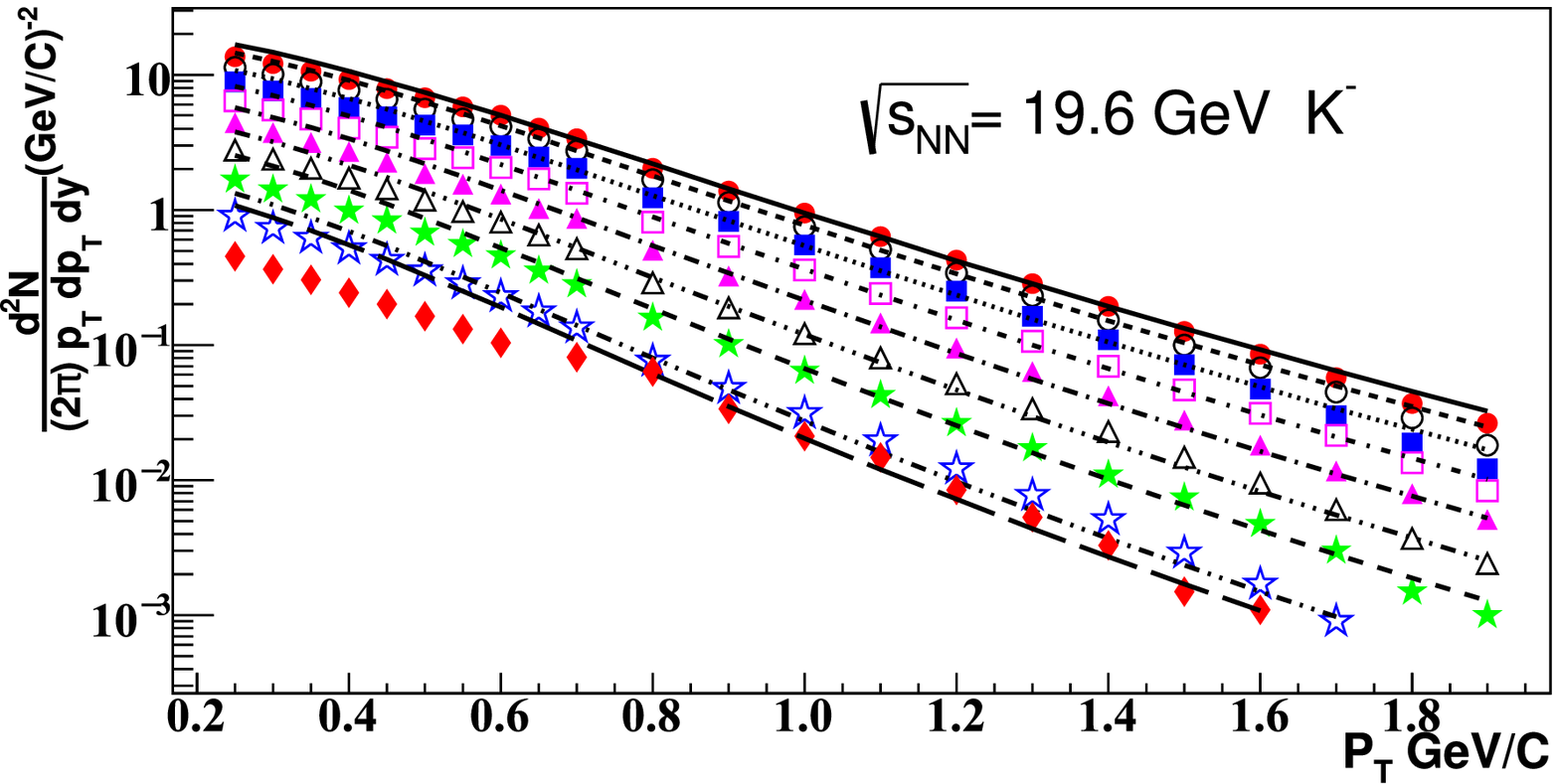}}\qquad  
   \subfigure[ ]{\label{fig:pip19_6}\includegraphics[height=5cm, width=7cm]{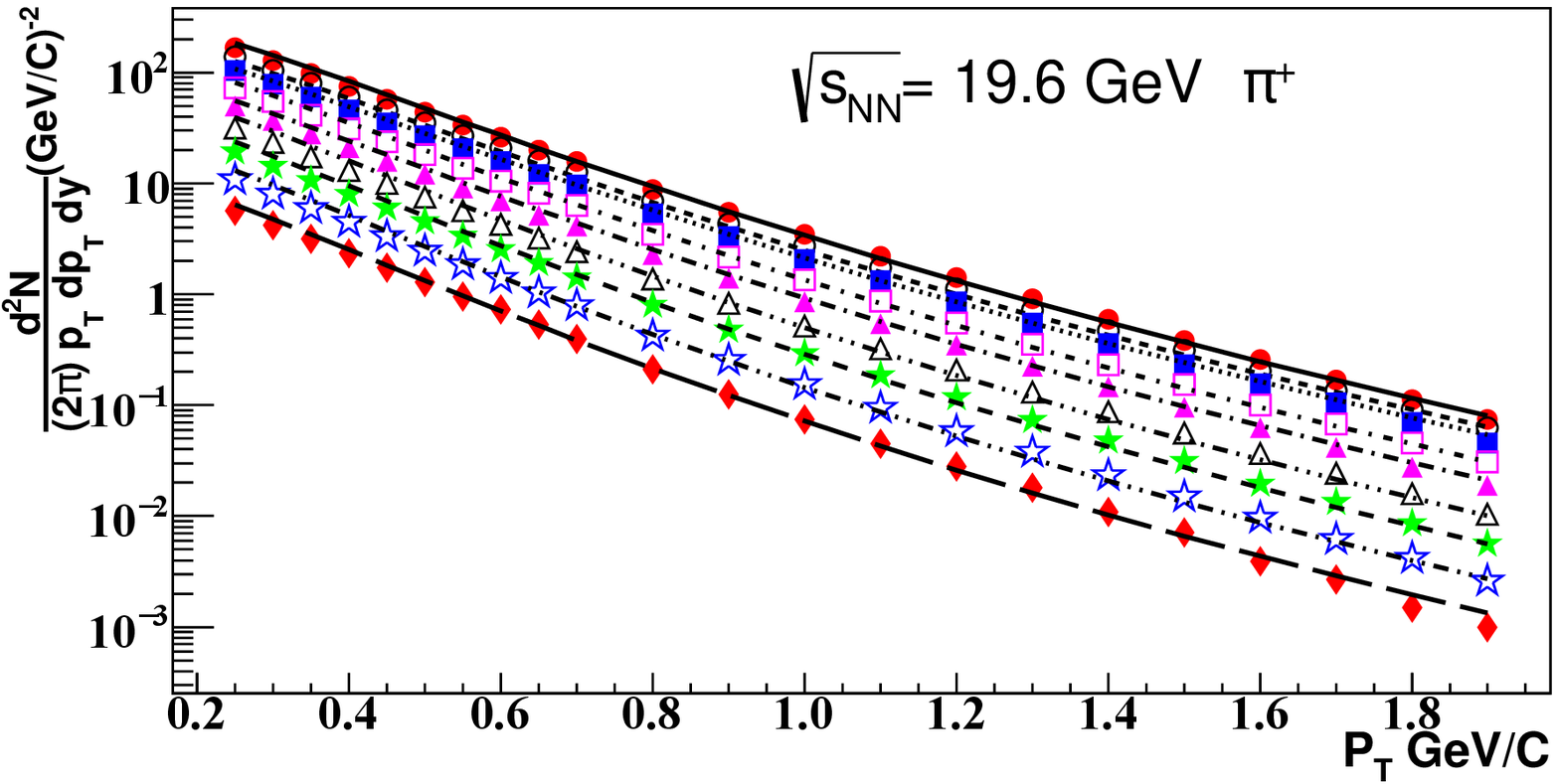}}\qquad  
 \subfigure[ ]{\label{fig:pim19_6}\includegraphics[height=5cm, width=7cm]{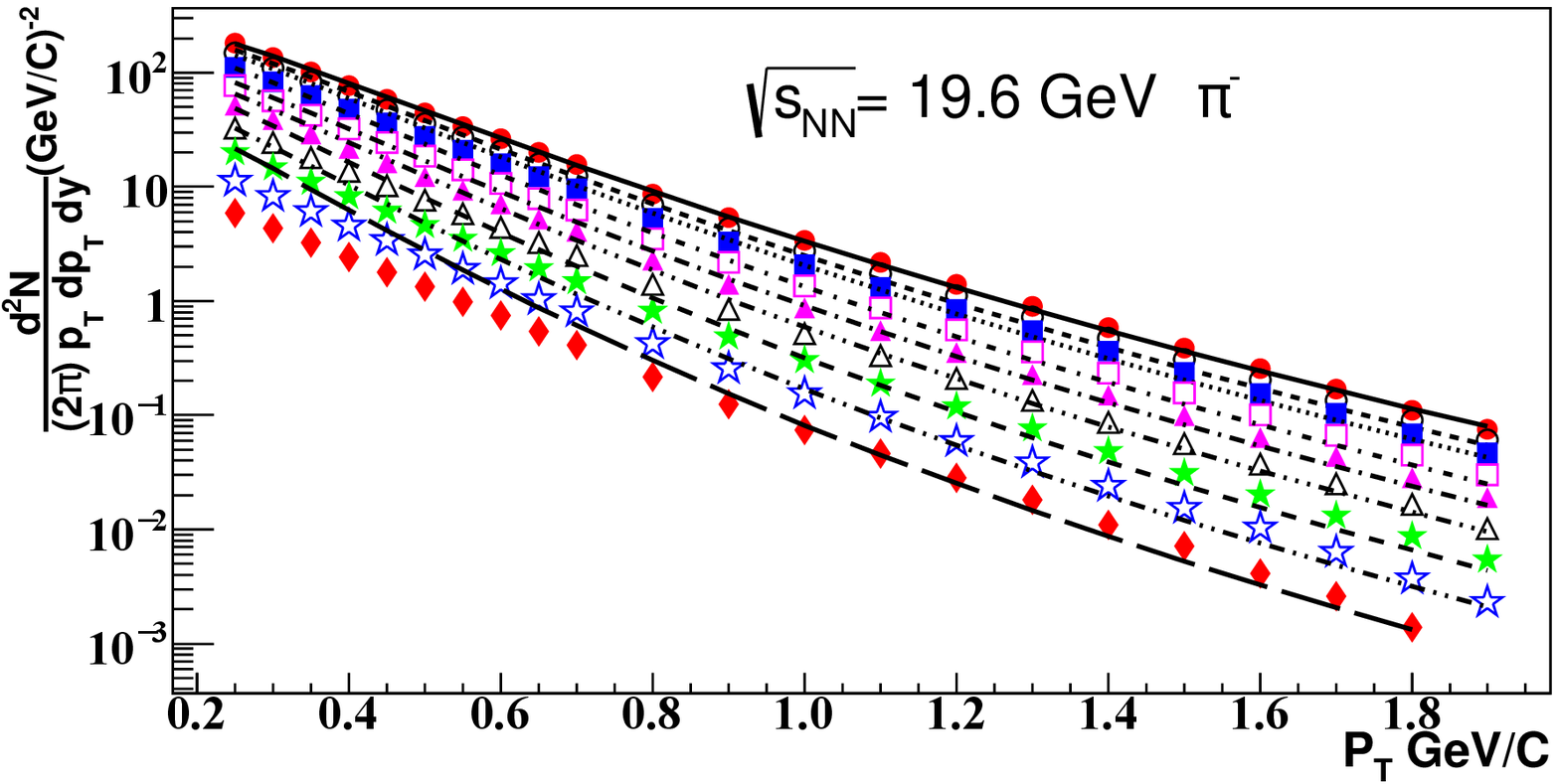}}\qquad 
 \subfigure[ ]{\label{fig:pbar19_6}\includegraphics[height=5cm, width=7cm]{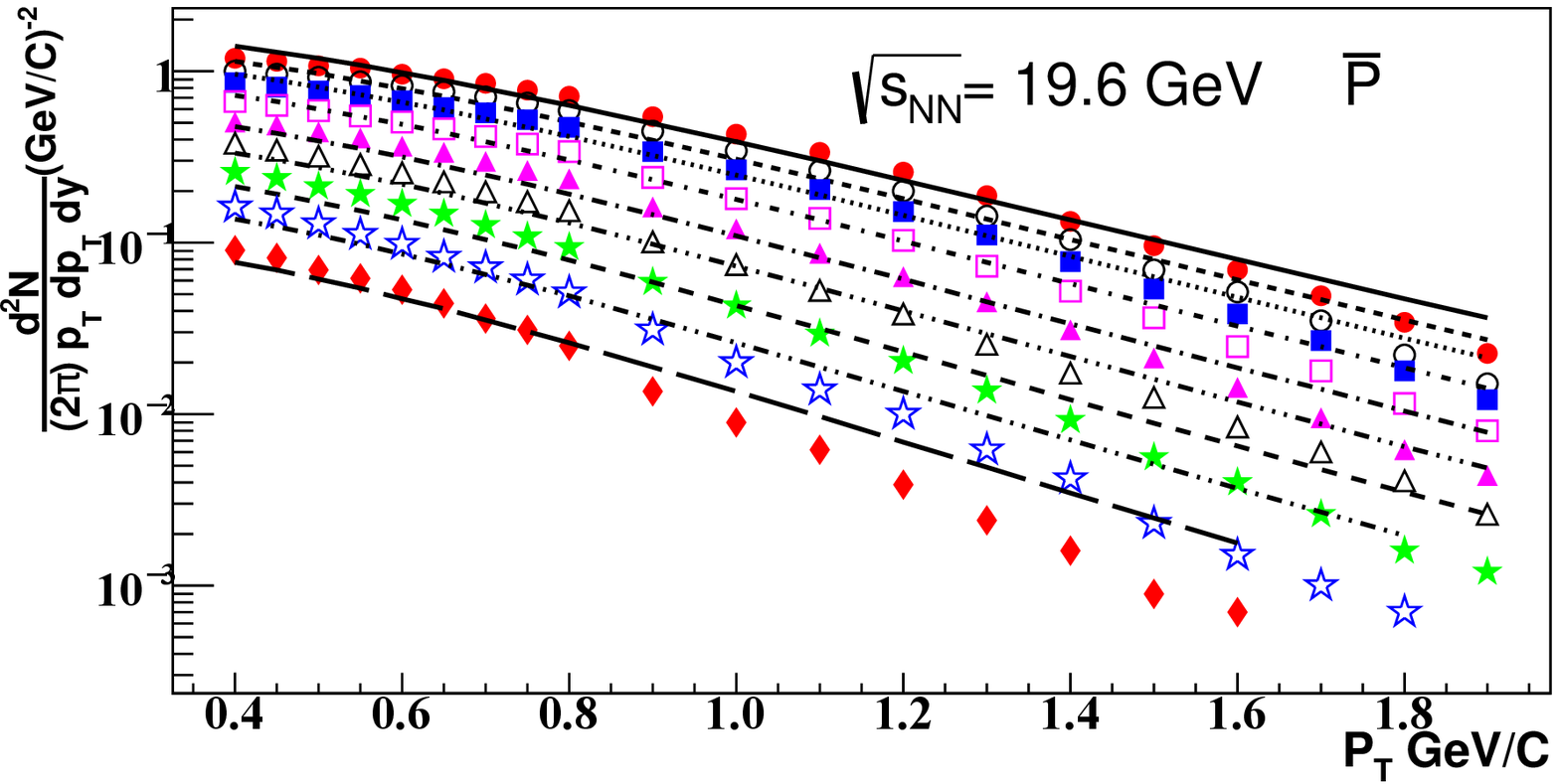}}\qquad  
 \subfigure[ ]{\label{fig:proton19_6}\includegraphics[height=5cm, width=7cm]{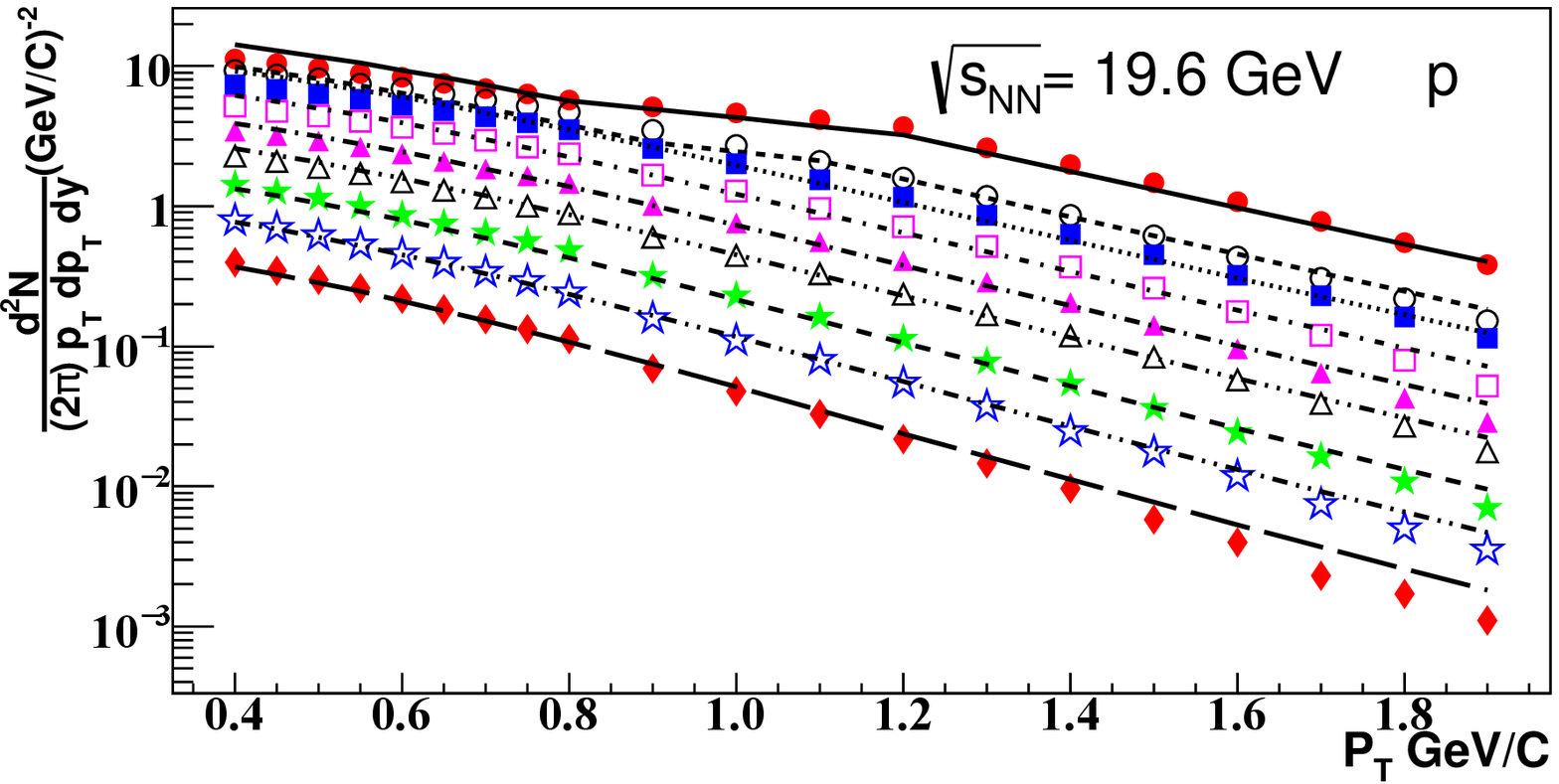}}\qquad 
 \caption[figtopcap]{The same as Fig (\ref{fig:7_7kaon}) at $\sqrt{s_{NN}} = 19.6 \hspace{0.05cm}GeV$ }
  \label{fig:19_6kaon}
\end{figure}

\begin{figure}[H] 
     \centering 
      \setlength\abovecaptionskip{-0.05\baselineskip} 
 \subfigure[ ]{\label{fig:kp27}\includegraphics[height=5cm, width=7cm]{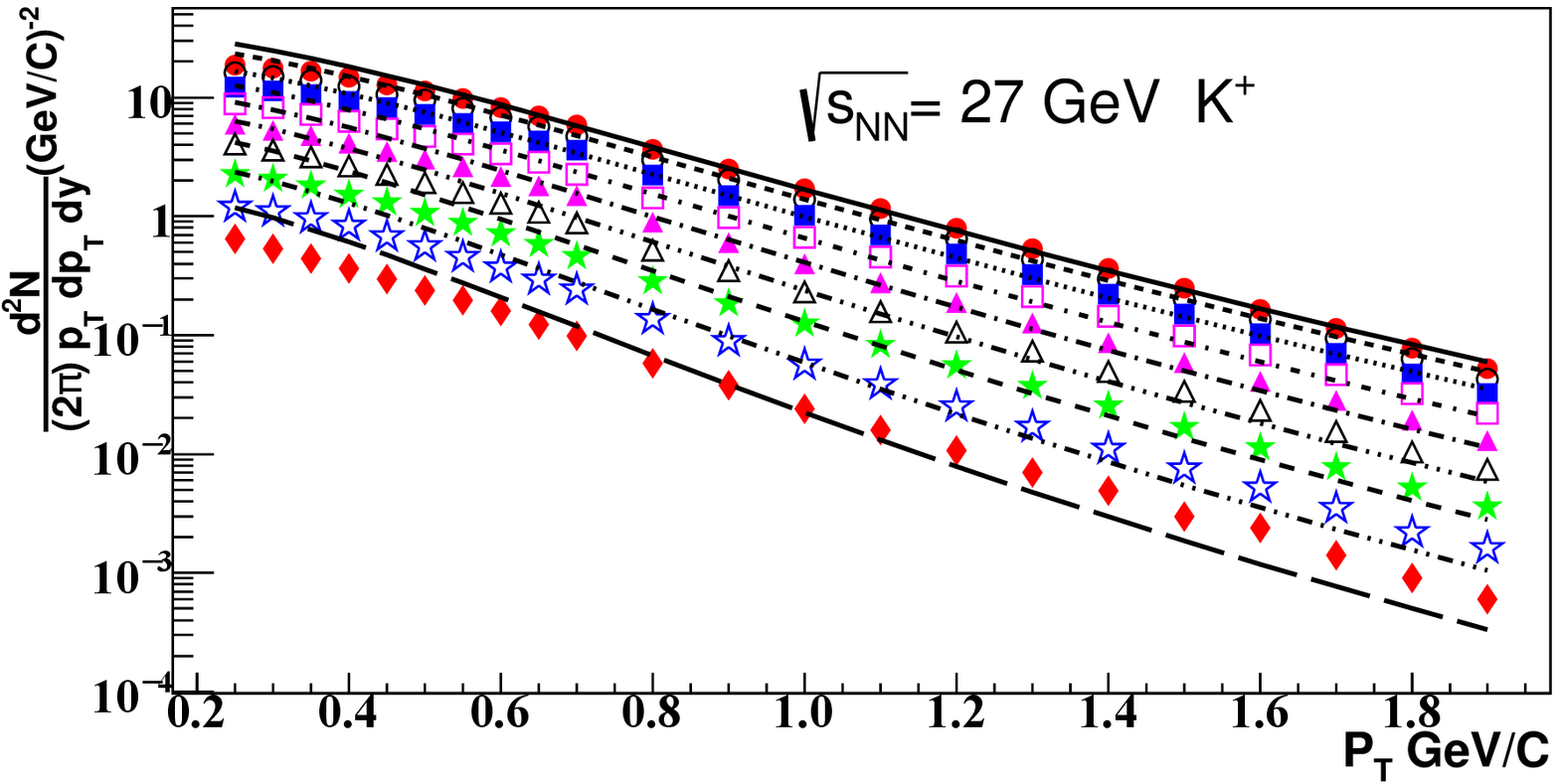}}\qquad
  \subfigure[ ]{\label{fig:km27}\includegraphics[height=5cm, width=7cm]{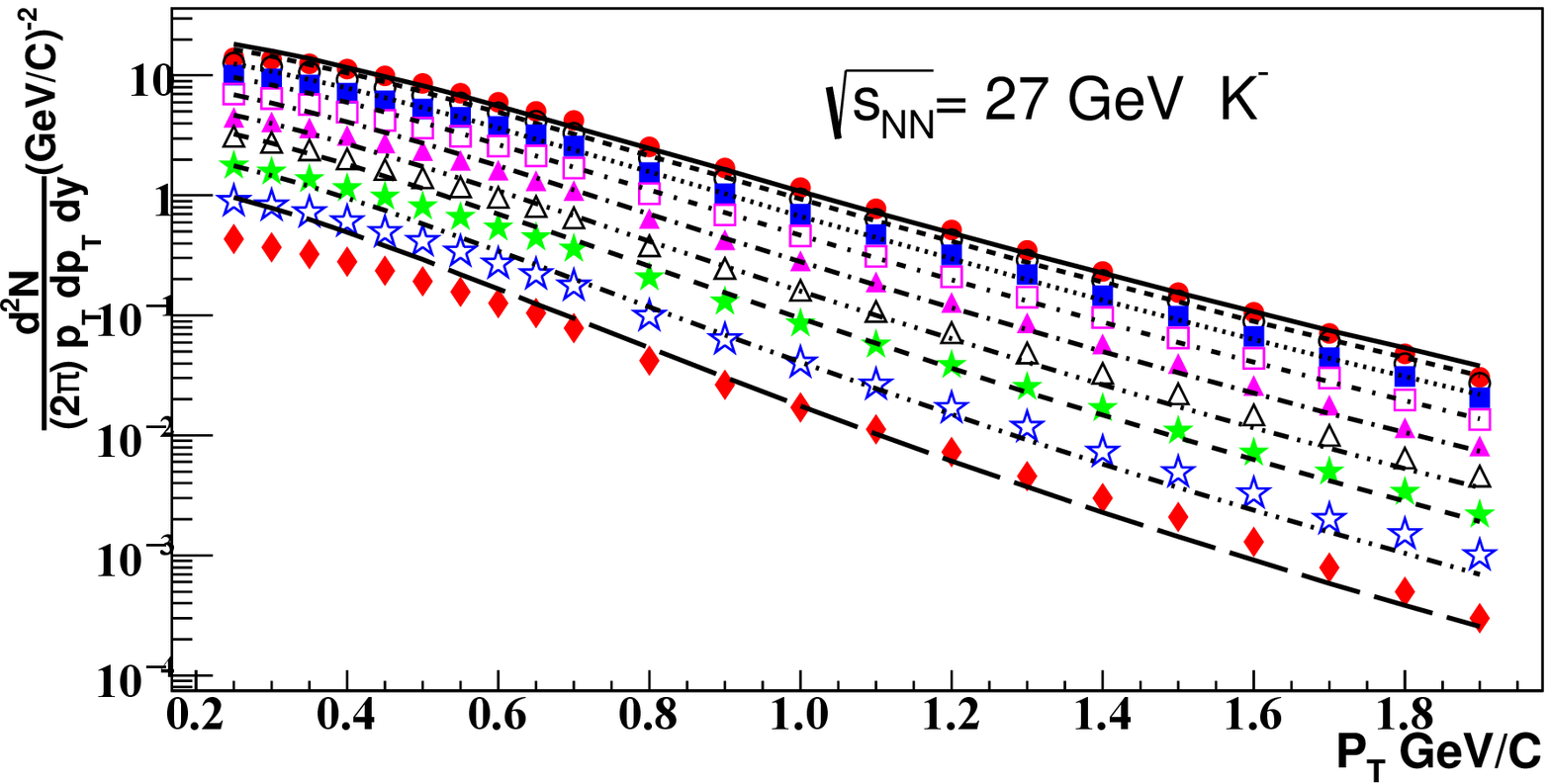}}\qquad  
   \subfigure[ ]{\label{fig:pip27}\includegraphics[height=5cm, width=7cm]{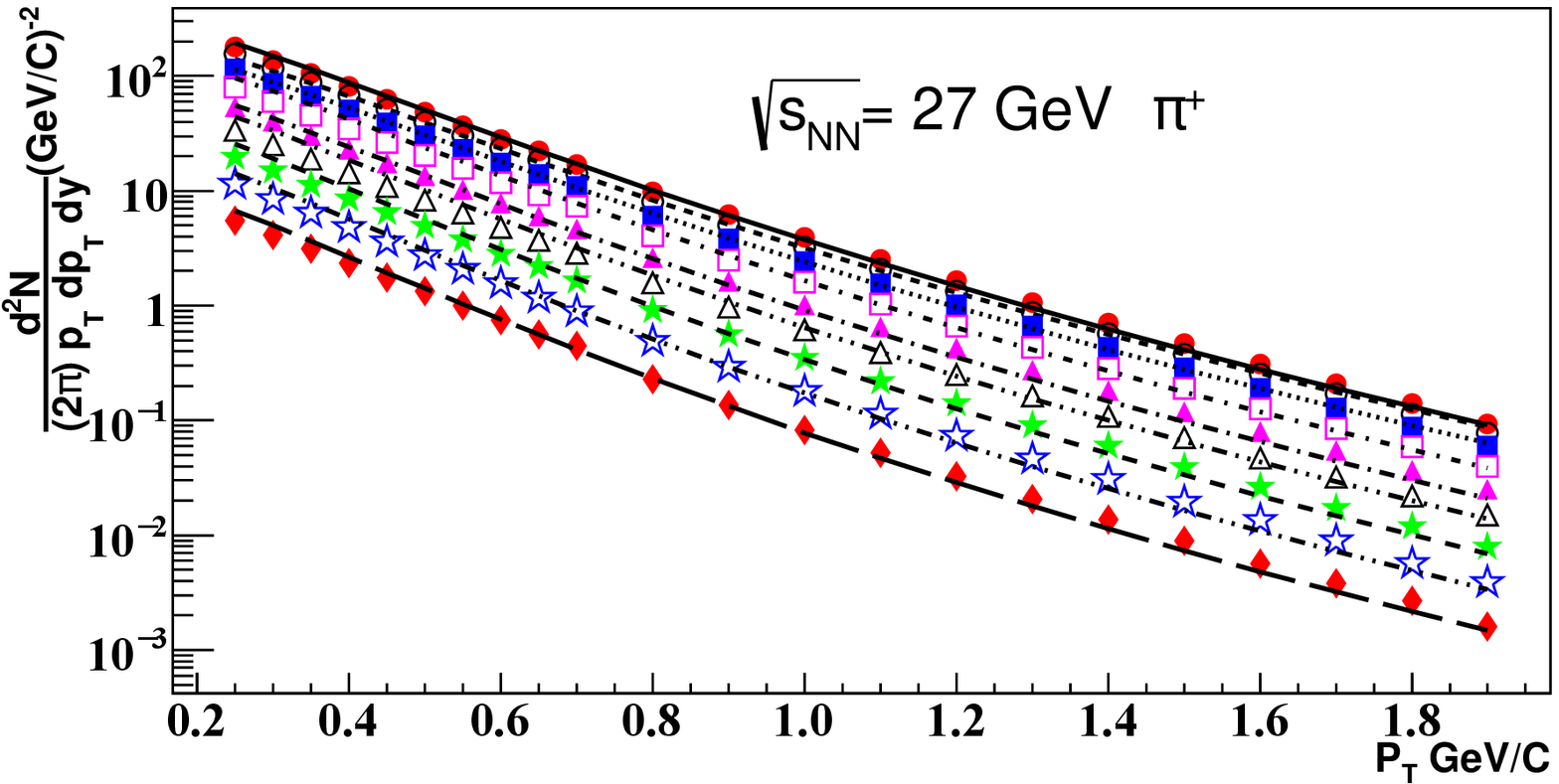}}\qquad  
 \subfigure[ ]{\label{fig:pim27}\includegraphics[height=5cm, width=7cm]{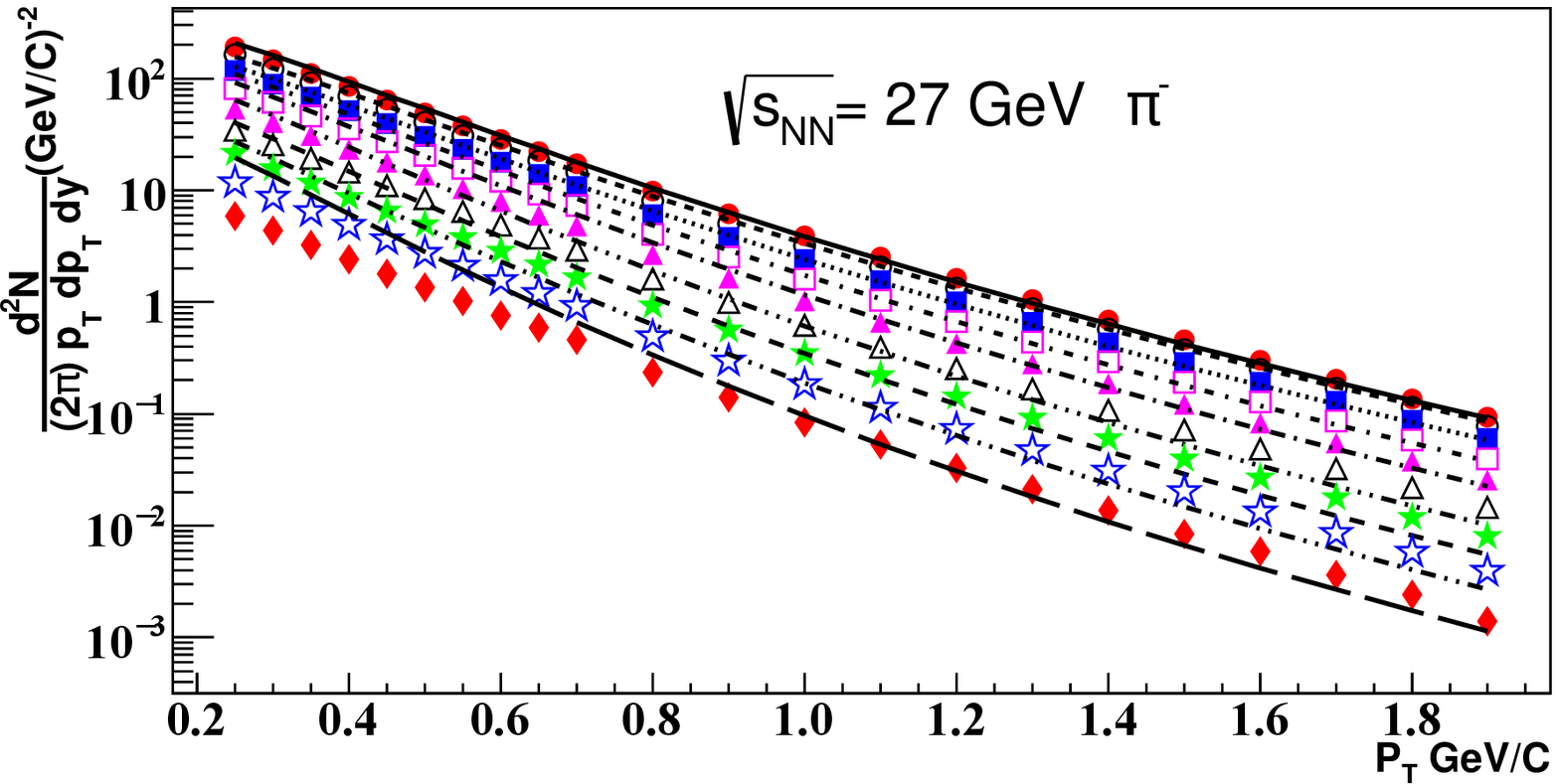}}\qquad 
 \subfigure[ ]{\label{fig:pbar27}\includegraphics[height=5cm, width=7cm]{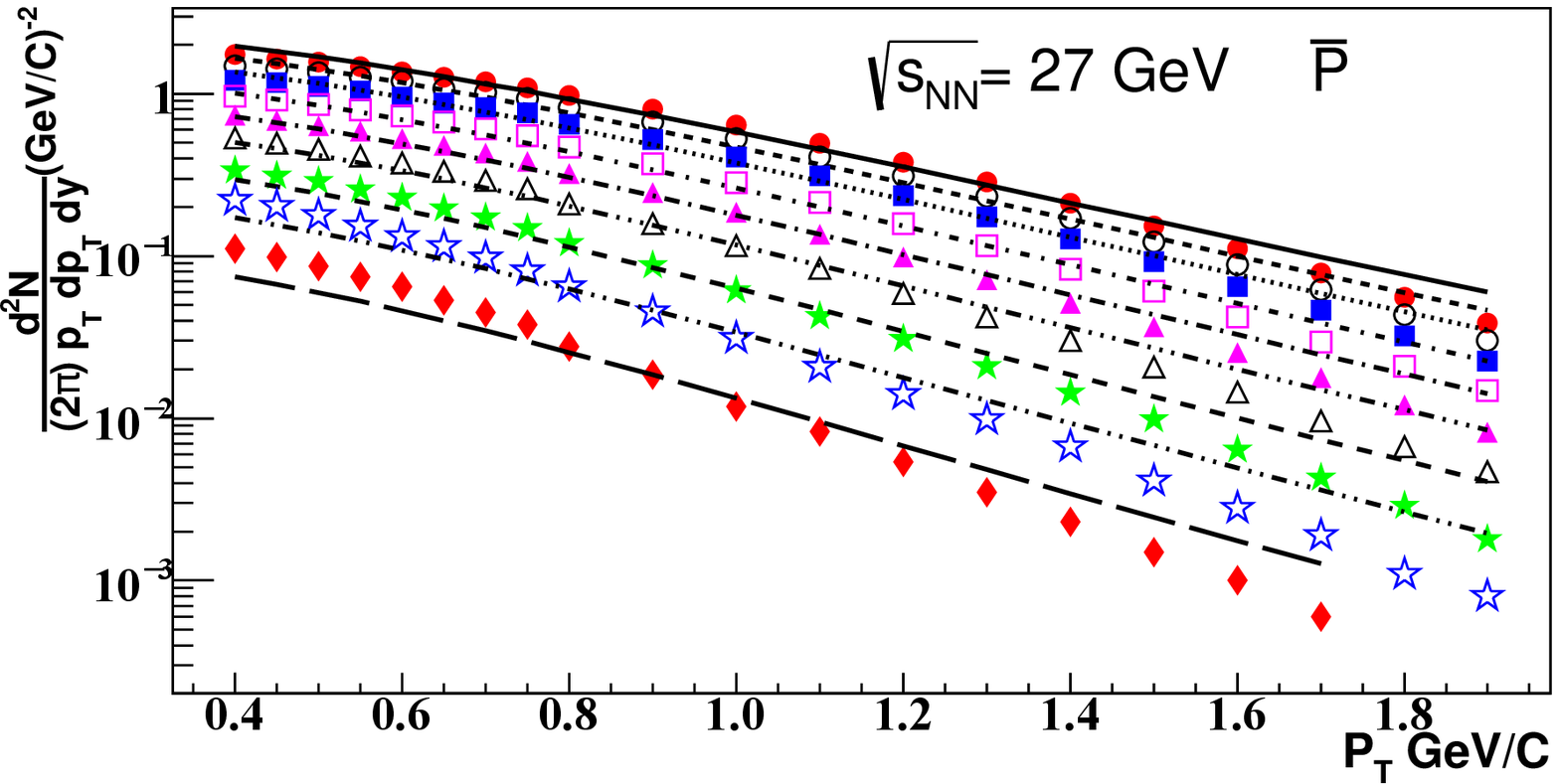}}\qquad  
 \subfigure[ ]{\label{fig:proton27}\includegraphics[height=5cm, width=7cm]{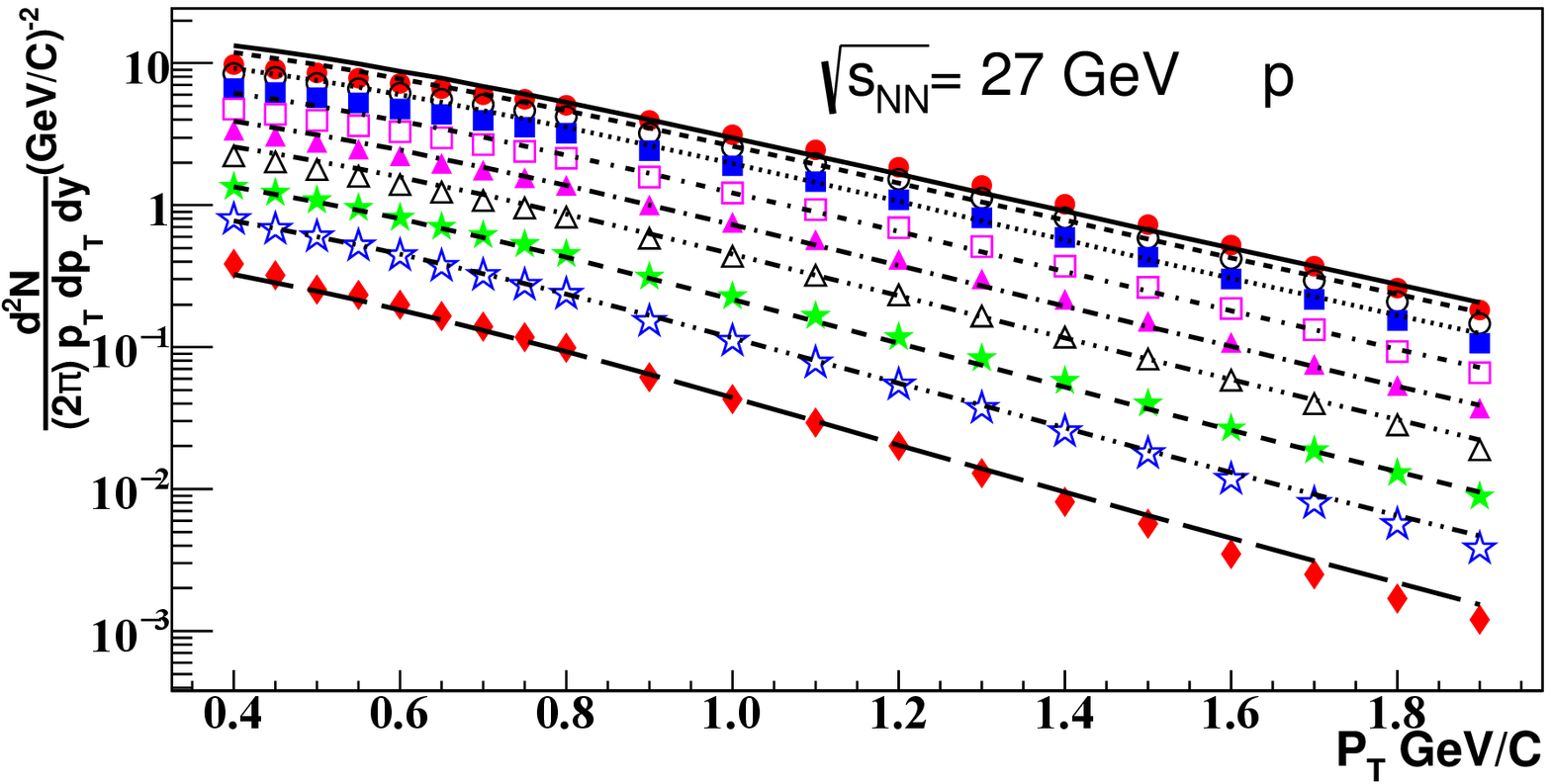}}\qquad 
 \caption[figtopcap]{ The same as Fig (\ref{fig:7_7kaon}) at $\sqrt{s_{NN}} = 27 \hspace{0.05cm}GeV$}
  \label{fig:27kaon}
\end{figure}

 The extracting parameters from fitting are: the temperature, T, the average volume which mainly was constant at the same energy, V, and the non-extensive parameter q. 
The parameters are listed in tables (\ref{table:kaon1}, \ref{table:pion2}, and \ref{table:proton3}). We have utilized $\chi^{2}$ to obtain the best fit with the experimental data, in addition we estimated the quadratic deviation $q^{2}$. They are defined as:

\begin{equation}
\chi^{2} = \sum_{i} \frac{\left( Y_{i}^{exp}-Y_{i}^{Theor} \right)^{2} }{\sigma_{i}^{2}}, \quad \hspace{1cm} q^{2} =  \sum_{i} \frac{\left( Y_{i}^{exp}-Y_{i}^{Theor} \right)^{2} }{\left( Y_{i}^{Theor}\right)^{2}}
\end{equation}
Where  $Y_{i}^{exp}$ is the transverse momentum distribution experimental value with the corresponding model calculations $Y_{i}^{Theor}$. $\sigma_{i}$ is the statistical error, and i refers to the particle species.

It is clear that by increasing the energy collision the $\chi^{2}/dof$ decreases an an example for the particle $k^{+}$ $\chi^{2}/dof \equiv 190.1/21 = 9.0529$ at energy from 7.7 GeV and then started to decrease to reach to 2.2 at 19.6 GeV. Then it starts to increase again.

This situation is repeated at different energies for the other particles, this is clearly appeared in the transverse momentum spectra when the energy increases. This may be attributed to that Tsallis distribution is suited very well at lower collision energy than at higher one.

\begin{table}
\caption{Extracted fitting parameters, T,V at ($ 0-5 \% $) for $k^{\pm}$, $V_{k^{\pm}}$ = (1.65, 1.81)$\times 10^6 \hspace{0.05cm}GeV^{-3}$ with  $\chi^2/dof$ and $q^{2}$ }
\begin{tabular}{|c| c| c| c| c|c|c|c|c|}
\hline 
 $\sqrt{s_{NN}}$(\hspace{0.05cm}GeV)  &Particle &T(\hspace{0.05cm}GeV) &  $\chi^{2}/dof$ & $q^{2}$ &Particle &T(\hspace{0.05cm}GeV) &  $\chi^2{}/dof$ & $q^{2}$ \\  
 \cline{1-9} 
 7.7  & $k^{+}$ &0.093 &  9.0529 &0.5855& $k^{-}$ &0.068 & 5.2140  & 1.32\\ 
 11.5 &         &0.095 &  3.8212 &0.3208&         &0.085 & 10.880  & 0.53\\
 19.6 &         &0.098 &  2.2641 &0.1442&         &0.097 & 2.7201  & 0.18\\
 27   &         &0.099 &  7.8619 &0.3487&         &0.099 & 2.3276  & 0.16\\
 39   &         &0.10  &  7.1408 &0.4086&         &0.1   & 5.4957  & 0.38\\ [1ex]
 \hline
\end{tabular}
\label{table:kaon1}
\end{table}

\begin{table}
\caption{The fitting parameter, T, at ($ 0-5 \% $) for $\pi^{\pm}$, $V_{\pi^{\pm}}$ = (2.99, 3.48)$\times 10^6 \hspace{0.05cm}GeV^{-3}$ with  $\chi^2/dof$ and $q^{2}$ }
\begin{tabular}{|c| c| c| c| c|c|c|c|c|}
\hline 
 $\sqrt{s_{NN}}$(\hspace{0.05cm}GeV)  &Particle &T(\hspace{0.05cm}GeV) &  $\chi^{2}/dof$ & $q^{2}$ &Particle &T(\hspace{0.05cm}GeV) &  $\chi^{2}/dof$ & $q^{2}$ \\  
 \cline{1-9} 
 7.7  &$\pi^{+}$ & 0.083 &  5.7464 & 0.39 & $\pi^{-}$ & 0.090  &  5.32996 &0.34 \\ 
 11.5 &          & 0.088 &  3.2168 & 0.27 &           & 0.080  &  3.21680 & 0.27 \\
 19.6 &          & 0.094 &  0.9972 & 0.08 &           & 0.094  &  0.9972  &0.08   \\
 27   &          & 0.095 &  0.9936 & 0.12 &           & 0.094  &  0.9936  &0.12   \\
 39   &          & 0.097 & 1.9460  & 0.24 &           & 0.096  &  1.9460  &0.24   \\ [1ex] 
 \hline
\end{tabular}
\label{table:pion2}
\end{table}
\begin{table}
\caption{The fitting parameter, T, at ($ 0-5 \% $) for $p,\overline{p}$, $V_{p,\overline{p}}$ = (5.56, 2.24)$\times 10^6 \hspace{0.05cm}GeV^{-3}$ with  $\chi^2/dof$ and $q^{2}$ }
\begin{tabular}{|c| c| c| c| c|c|c|c|c|}
\hline 
 $\sqrt{s_{NN}}$(\hspace{0.05cm}GeV)  &Particle &T(\hspace{0.05cm}GeV) &  $\chi^{2}/dof$ & $q^{2}$ &Particle &T(\hspace{0.05cm}GeV) &  $\chi^{2}/dof$ & $q^{2}$ \\  
 \cline{1-9} 
 7.7  & p & 0.097  & 4.5909 & 0.49 &$\overline{p}$& 0.13  & 0.4330  & 0.07\\ 
 11.5 &   & 0.093  & 3.3253 & 0.47 &              & 0.08  & 0.87779 & 0.96\\
 19.6 &   & 0.108  & 1.8245 & 0.26 &              & 0.135 & 1.56922 & 0.26\\
 27   &   & 0.107  & 3.2119 & 0.38 &              & 0.145 & 2.4239  & 0.34\\
 39   &   & 0.109  & 6.2037 & 0.64 &              & 0.154 & 2.9908  & 0.41\\ [1ex] 
 \hline
\end{tabular}
\label{table:proton3}
\end{table}

\begin{figure}[H] 
     \centering 
      \setlength\abovecaptionskip{-0.05\baselineskip} 
  \includegraphics[height=7cm, width=10cm]{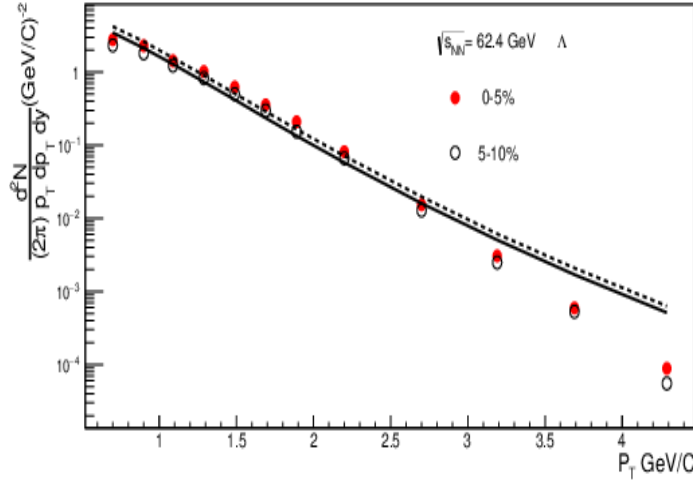}\qquad
 \caption[figtopcap]{The Colored-online , solid and dashed lines represent the transverse momentum distribution $ d^{2}N/2\pi P_{T} dP_{T} dy $ versus the transverse momentum $p_{T}$ for $ \Lambda $ particles in Au+Au collisions at $\sqrt{s_{NN}} = 62.4 \hspace{0.05cm}GeV$ for $ 0-5 \%$ and $ 5-10 \%$ . The Colored-online symbols are the experimental data \cite{aggarwal2011strange}}
  \label{fig:62_4kaon}
\end{figure}

In Fig (\ref{fig:62_4kaon}) we show the $p_{T}$ $-$spectra again for $ \Lambda $ particle at $\sqrt{s_{NN}} = 62.4 \hspace{0.05cm}GeV $ for the centerality classes $ 0-5 \% $ and $ 5-10 \% $ \cite{zheng2015can,aggarwal2011strange}.
We find a good fitting results especially in low $ P_{T}$ starting from $  0.50\gtrsim p_{T}\gtrsim 2.50 GeV/C $, while the distribution deviates at higher $ p_{T}$. The extracted fitting parameters are listed in table (\ref{table:4lmd})

\begin{table}[h!]
\centering
\begin{tabular}{c |c |c| c|c} 
 \hline
 $\sqrt{s_{NN}} = 64.2 $(\hspace{0.05cm}$GeV$)  & T(\hspace{0.05cm}GeV) & V  & $\chi^2/dof$ & $q^{2}$\\ [0.5ex] 
 \hline
 $ 0-5 \%$   & 0.11 & 4.9$\times 10^6$ & 8.0/12& 0.011\\ [1ex] 
 $ 5-10 \%$  & 0.11 & 3.9$\times 10^6$ & 3.2/12& 0.007\\ [1ex] 
 \hline
\end{tabular}
\caption{Extracted fitting parameters $T$ and $V$ for $ \Lambda $ at centrality  $ 0-5 \% $, and $ 5-10 \% $.}
\label{table:4lmd}
\end{table}

By using tables (\ref{table:kaon1}, \ref{table:pion2}, \ref{table:proton3}), we have fitted the temperature and the center of mass energy for all particles for the selected centrality $(0-5\%)$ eq.(\ref{eq:Tlim1}). The fitted relation is written as:

\begin{equation} \label{eq:Tlim1}
T _{(0-5\%)} [GeV] = a \hspace{0.07cm} ln( \sqrt{s_{NN}} ) + b
\end{equation}

Where the values of $ a $ and $ b $ are attached to the particles as listed in the following Table \ref{table:fit_eq}:

\begin{table}[h!]
\centering
\begin{tabular}{|c| c| c|} 
 \hline
 Particle & a (GeV) & b (GeV)  \\ [0.5ex] 
 \hline\hline
 $k^{+}$        & 0.0044   & 0.0848  \\ 
 $k^{-}$        & 0.0129   & 0.0351   \\
 $\pi^{+}$      & 0.0089   & 0.0665    \\
 $\pi^{-}$      & 0.007    & 0.071      \\
 $p $           &  0.009   & 0.0771      \\  
 $\overline{p}$ & 0.0284   & 0.0473       \\ [1ex]
 \hline
\end{tabular}
\caption{Values of $ a $ and $ b $ parameters correspond to each particle}
\label{table:fit_eq}
\end{table}

In order to make more investigation and explore the applicability of Tsallis distribution, we have applied eq.(\ref{eq:num}) in order to calculate the ratios of $k^{+}/\pi^{+}$, and $\Lambda/\pi^{-}$
\begin{figure}[H] 
     \centering 
      \setlength\abovecaptionskip{-0.05\baselineskip} 
 \subfigure[ ]{\label{fig:kp_pip1}\includegraphics[height=5.5cm, width=7cm]{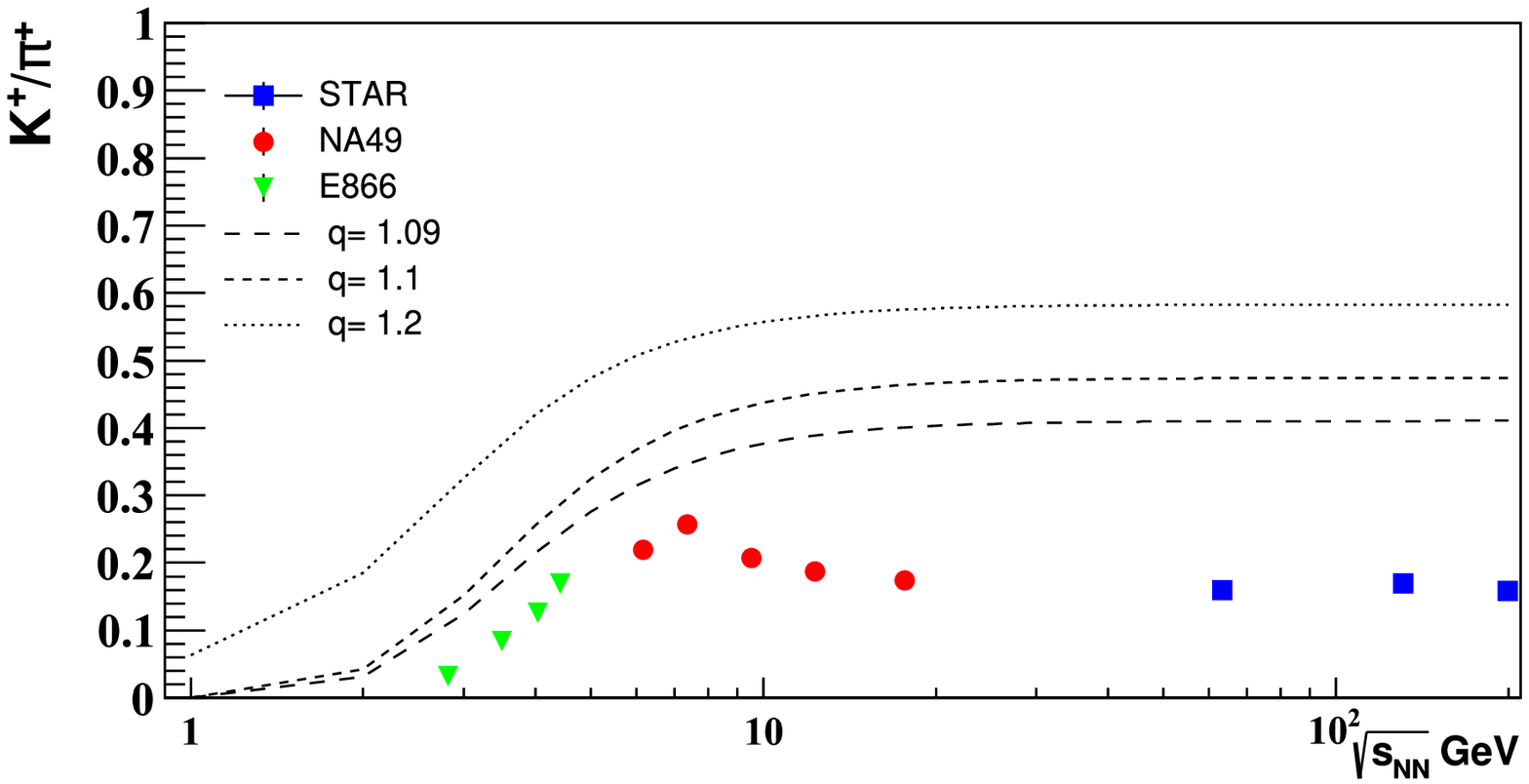}}\qquad
  \subfigure[ ]{\label{fig:lam_pim}\includegraphics[height=5.5cm, width=7cm]{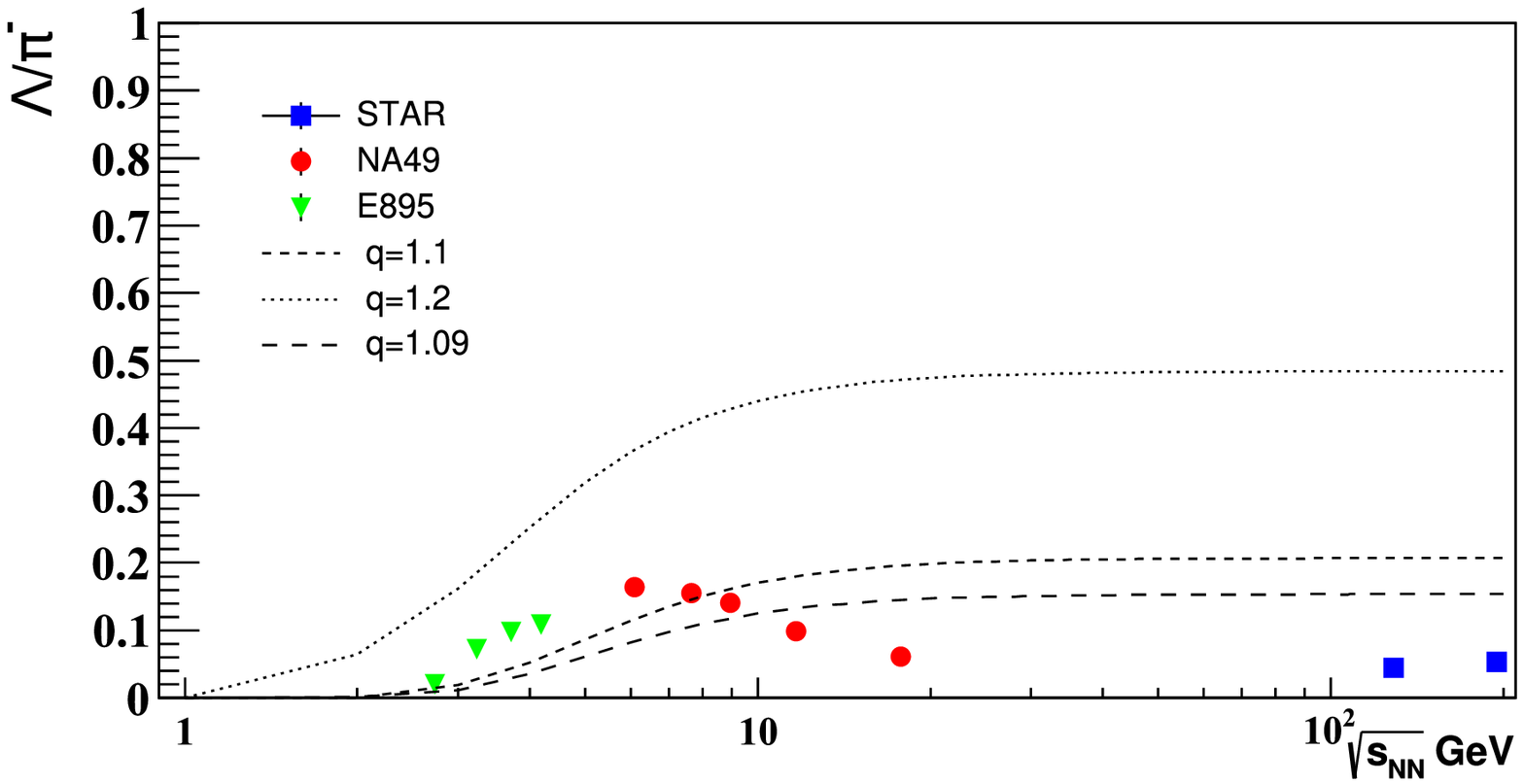}}\qquad
  \caption[figtopcap]{The particle ratios versus collision energy $\sqrt{s_{NN}}$ \subref{fig:kp_pip1} and \subref{fig:lam_pim}, both are calculated at q = 1.1. The Colored-online symbols are the experimental data for    $ \frac{k^{+}}{\pi^{+}}$ \cite{bugaev2013simple,adams2004identified,afanasiev2002energy,alt2008pion}, and for $\frac{\Lambda}{\pi_{-}}$ \cite{klay2003charged,anticic2004lambda,pinkenburg2001production,alt2008energy}}
  \label{fig:ratio}
 \end{figure}
  
The ratios of particles are represented in Fig.(\ref{fig:ratio}) such as (Fig. $k^{+}/\pi^{+} $ (\ref{fig:kp_pip1}), and Fig. $ \Lambda/ \pi^{-} $ (\ref{fig:lam_pim} ). The number density of each particle at zero chemical potential is calculated by utilizing eq.(\ref{eq:num}) with the nonextensive parameter , $q = 1.1$, and the temperature expression is taken as \cite{andronic2006hadron},
 
\begin{equation} \label{eq:temp}
T[MeV] = T_{lim} \hspace{0.05cm} \left[1-\frac{1}{0.7+(exp(\sqrt{S_{NN}}-2.9))/1.5} \right] 
\end{equation}
Where $\sqrt{S_{NN}}$ in GeV, $T_{lim} = \left( 161 \pm 4\right)  \hspace{0.03cm} MeV$.

The ratios are confronted with the experimental data which taken from E895 \cite{klay2003charged, pinkenburg2001production}, NA49 \cite{afanasiev2002energy,alt2008pion,anticic2004lambda,alt2008energy} and STAR \cite{adams2004identified}.

It is clear that, the ratio ($k^{+}/\pi^{+} $) as shown in fig. \ref{fig:kp_pip1} agrees roughly well with the experimental data only at low energy (i.e $\sqrt{S_{NN}} < 10 \hspace{0.05cm}GeV$), but overestimate and slightly increases as the energy increases for $\sqrt{S_{NN}} > 10 \hspace{0.05cm}GeV$. Additionally, it is worth to shed light on the horn of the ratio that appears clearly in the experimental data and in the thermal model \cite{andronic2006hadron,tawfik2016particle,bugaev2013simple} based on the extensive thermodynamics, but disappears in the non-extensive model. The disappearance of the horn also occurs in ($\Lambda/ \pi^{-} $) ratio as shown in fig. \ref{fig:lam_pim}. Additionally, the ratios are repeated at different values of the entropic index ($q = 1.09,1.1, 1.2$). We notice that by increasing q the model ratios deviate more aparted from the experminatl data even at low energy. Accordingly, we assure that this behavior of the non-extensive ratios does not mean a solution of the horn puzzle.   

\section{Conclusion}
 In the present work, the transverse momentum spectra of some mesons such as (pion and kaon) and some baryons such as (p, $\overline{p}$) produced in Au–Au collisions at $\sqrt{S_{NN}}=7.7, 11.5, 19.6, 27, 39 \hspace{0.05cm}GeV$, and $\Lambda$ at 62.4 GeV have been studied based on Tsallis statistics. Characterization of each identified particle is studied elaborating $\chi^2$ fit with the experimental data at different centralities and at mid-rapidity ($y = 0$) and zero chemical potential. A good fit is noted for pions and $\Lambda$. The latter is the best where $\chi^2/12$ is 0.66 for $ 0-5 \% $  and 0.26 for $ 5-10 \% $.
Finally, we conclude that the Tsallis distribution, gives an a clear and satisfied successful description of the $P_{T}$ spectra for AA collision at BES. Moreover, Tsallis distribution explains the particle ratio fairly well at only small energy but dramatically deviates at large energy, in addition to that it does not exhibit a horn but slightly continuous flat.
The values of the Tsallis parameter, q, is fixed for all particles under estimation to be in the range 1.09. However, we expect it can be increased with increasing the energy. The volume also is extracted for each particle at each energy but it slightly changed for group of particles such as pions, kaons, and proton, antiproton.

\section{References}
\providecommand{\href}[2]{#2}\begingroup\raggedright
  
\end{document}